\documentclass[twocolumn,english,aps,prl,superscriptaddress,showpacs]{revtex4}
\usepackage[T1]{fontenc}
\usepackage[latin9]{inputenc}
\usepackage{bm}
\usepackage{amsmath}
\usepackage{amssymb}
\usepackage{esint}

\makeatletter
\@ifundefined{textcolor}{}
{%
 \definecolor{BLACK}{gray}{0}
 \definecolor{WHITE}{gray}{1}
 \definecolor{RED}{rgb}{1,0,0}
 \definecolor{GREEN}{rgb}{0,1,0}
 \definecolor{BLUE}{rgb}{0,0,1}
 \definecolor{CYAN}{cmyk}{1,0,0,0}
 \definecolor{MAGENTA}{cmyk}{0,1,0,0}
 \definecolor{YELLOW}{cmyk}{0,0,1,0}
}

\usepackage{bm}

\makeatother

\usepackage{babel}
\begin{document}

\title{Energy Magnetization and Thermal Hall Effect}

\author{Tao Qin}

\affiliation{Institute of Physics, Chinese Academy of Sciences, Beijing 100190,
China}

\author{Qian Niu }

\affiliation{Department of Physics, The University of Texas at Austin, Austin,
Texas 78712, USA}

\affiliation{International Center for Quantum Materials, Peking University, Beijing
100871, China}

\author{Junren Shi}

\email{junrenshi@pku.edu.cn}

\affiliation{International Center for Quantum Materials, Peking University, Beijing
100871, China}
\begin{abstract}
We obtain a set of general formulae for determining magnetizations,
including the usual electromagnetic magnetization as well as the gravitomagnetic
energy magnetization. The magnetization corrections to the thermal
transport coefficients are explicitly demonstrated. Our theory provides
a systematic approach for properly evaluating the thermal transport
coefficients of magnetic systems, eliminating the unphysical divergence
from the direct application of the Kubo formula. For a non-interacting
anomalous Hall system, the corrected thermal Hall conductivity obeys
the Wiedemann-Franz law.
\end{abstract}


\maketitle
Thermal Hall effect, or Righi-Leduc effect, is the thermal analogue
of the Hall effect~\cite{deGroot1984}. It gives rise to a transverse
heat flow when a temperature gradient is applied. There are recent
surging experimental interests in studying the thermal Hall effect
in various systems, revealing such as the phonon Hall effect~\cite{Strohm2005,Inyushkin2007},
magnon Hall effect~\cite{Onose2010}, and so on. There are also many
theoretical efforts to study these phenomena~\cite{Sheng2006,Kagan2008,Zhang2010,Katsura2010}.
However, most of these theoretical studies face a fundamental issue:
direct application of the Kubo formula, when done correctly without
questionable ``tricks''~%
\footnote{A common trick used in the bosonic calculation is to ignore the terms
involving $aa$ and $a^{\dagger}a^{\dagger}$ in the energy current
operator, where $a$ ($a^{\dagger}$) is the bosonic annihilation
(creation) operator, as pointed out in Ref~\cite{Zhang2010}. This
will result in an apparently converging albeit incorrect thermal Hall
coefficient. %
}, always yields unphysical divergence at the zero temperature~\cite{Agarwalla2011,Matsumoto2011}.
Such an issue is actually a major obstacle in the theoretical studies
of the thermal Hall effect.

The underlying reason of the issue had been previously identified~\cite{Smrcka1977,Cooper1997}:
in a system breaking the time-reversal symmetry, either by applying
an external magnetic field or due to the spontaneous magnetization,
the temperature gradient not only drives the transport (heat) current,
but also drives the circulating (heat) current that is not observable
in the transport experiment. Both contributions are present in the
microscopic current density calculated by the standard linear response
theory, and a proper subtraction of non-observable circulating component
is necessary. For the electric transport, such subtraction involves
the electromagnetic orbital magnetization density, while the subtraction
of the energy current will involve the garvitomagnetic energy magnetization
density~\cite{Ryu2010}, which characterizes the circulating energy
flow. However, the previous theoretical discussions do not clarify
what the transport current and the magnetizations are, and how the
magnetizations can be evaluated for a general extended system. The
issue becomes more fundamental because the magnetizations are gauge-dependent
quantities~\cite{Hirst1997}, and it is not \emph{a-priori} clear
what the proper gauges of the magnetizations should be when calculating
the transport coefficients .

In this Letter, we attempt to build the theory of thermal transport
of magnetic systems on a firmer basis. We obtain a set of general
formulae for determining the magnetizations, including the usual electromagnetic
orbital magnetization as well as the gravitomagnetic energy magnetization
{[}Eqs.~(\ref{eq:MN-mu}--\ref{eq:MQ-T}){]}. We further show that
these magnetizations naturally emerge as corrections to the thermal
transport coefficients, recovering the Onsager relations and Einstein
relations {[}Eq.~(\ref{eq:Linear response}){]}, and eliminating
the unphysical divergence. The result is a complete set of general
formulae for calculating the transport thermal Hall conductivity,
as well as the other thermal-electric responses such as Nernst effect
and Ettingshausen effect~\cite{Xiao2006}. The formula also clarify
what the gravitomagnetic energy magnetization is and how it can be
calculated, and its thermodynamics is determined. We test our theory
by calculating the thermal Hall coefficient of a non-interacting anomalous
Hall system, and observe the emergence of Wiedemann-Franz law, consistent
to the recent experimental observation~\cite{Onose2008}.

\medskip{}

\noindent \emph{Preliminaries:} To make our discussion specific, we
consider a general electronic system. We should note that the formulae
we will develop are general, applicable to the other systems such
as the phonon and spin systems. 

We assume that the total Hamiltonian of the unperturbed system can
be written as $\hat{H}=\int\mathrm{d}\bm{r}\hat{h}(\bm{r})$, where
$\bm{r}$ denotes the spatial coordinate, and $\hat{h}(\bm{r})$ is
the local energy density operator. To study the electric and thermal
responses, we introduce the external mechanic fields: the potential
$\phi(\bm{r})$ and the gravitational field $\psi(\bm{r})$, where
the gravitational field is introduced as the mechanic counterpart
of the temperature gradient, following Luttinger~\cite{Luttinger1964}.
In the presence of these fields, the local energy density operator
of the system is modified to~\cite{Cooper1997}:
\begin{equation}
\hat{h}_{\phi,\psi}(\bm{r})=\left[1+\psi(\bm{r})\right]\left[\hat{h}(\bm{r})+\phi(\bm{r})\hat{n}(\bm{r})\right]\,,\label{eq:H_phi_psi}
\end{equation}
where $\hat{n}(\bm{r})$ is the local density operator, and the Hamiltonian
of the system is $\hat{H}_{\phi,\psi}=\int\mathrm{d}\bm{r}\hat{h}_{\phi,\psi}(\bm{r})$.

The particle and energy current operators of the system are defined
by the conservation equations~\cite{Cooper1997}:
\begin{align}
\frac{\partial\hat{n}(\bm{r})}{\partial t} & \equiv\frac{1}{\mathrm{i}\hbar}[\hat{n}(\bm{r}),\hat{H}_{\phi,\psi}]=-\bm{\nabla}\cdot\hat{\bm{J}}_{N}^{\phi,\psi}(\bm{r})\,,\label{eq:charge-continuity}\\
\frac{\partial\hat{h}^{\phi,\psi}(\bm{r})}{\partial t} & \equiv\frac{1}{\mathrm{i}\hbar}[\hat{h}^{\phi,\psi}(\bm{r}),\hat{H}_{\phi,\psi}]=-\bm{\nabla}\cdot\hat{\bm{J}}_{E}^{\phi,\psi}(\bm{r})\,,\label{eq:energy-continuity}
\end{align}
where $\hat{\bm{J}}_{N}^{\phi,\psi}$ and $\hat{\bm{J}}_{E}^{\phi,\psi}$
are particle and energy current operators, respectively. 

We further require that the current operators in the presence of the
external fields can be related to the zero-field current operators
$\hat{\bm{J}}_{N}$ and $\hat{\bm{J}}_{E}$ by~\cite{Cooper1997}:
\begin{align}
\hat{\bm{J}}_{N}^{\phi,\psi}(\bm{r}) & =\left[1+\psi(\bm{r})\right]\hat{\bm{J}}_{N}(\bm{r})\,,\label{eq:Scaling-N}\\
\hat{\bm{J}}_{E}^{\phi,\psi}(\bm{r}) & =\left[1+\psi(\bm{r})\right]^{2}\left[\hat{\bm{J}}_{E}(\bm{r})+\phi(\bm{r})\hat{\bm{J}}_{N}(\bm{r})\right]\,.\label{eq:Scaling-E}
\end{align}
We note that the current operator is only defined up to a curl by
Eqs.~(\ref{eq:charge-continuity}--\ref{eq:energy-continuity}).
As we will show later {[}See Eq.~(\ref{eq:JEproper}){]}, one may
use this freedom to find appropriate forms of current operators that
do satisfy these scaling relations~%
\footnote{The deviations from the scaling laws of the second or higher order
gradients of $\phi$ and $\psi$ will not affect our results. The
first gradient deviation can be eliminated by redefining the current
operators, as shown in Eq.~(\ref{eq:JEproper}).%
}.

When the system is in equilibrium and in the absence of the external
fields, we have $\bm{\nabla}\cdot\bm{J}_{N}^{\mathrm{eq}}=\bm{\nabla}\cdot\bm{J}_{E}^{\mathrm{eq}}=0$,
where $\bm{J}_{N(E)}^{\mathrm{eq}}(\bm{r})=\left\langle \hat{\bm{J}}_{N(E)}(\bm{r})\right\rangle _{0}$
is the expectation value of the particle (energy) current for the
equilibrium density matrix $\hat{\rho}_{0}=(1/Z_{0})\exp\left[-\hat{K}/k_{B}T_{0}\right]$,
where $\hat{K}=\int\mathrm{d}\bm{r}\hat{K}(\bm{r})$ and $\hat{K}(\bm{r})\equiv\hat{h}(\bm{r})-\mu_{0}\hat{n}(\bm{r})$.
As a result, we can introduce the zero field particle magnetization
density $\bm{M}_{N}(\bm{r})$ and the energy magnetization density
$\bm{M}_{E}(\bm{r})$ so that: 
\begin{equation}
\bm{J}_{N(E)}^{\mathrm{eq}}(\bm{r})=\bm{\nabla}\times\bm{M}_{N(E)}(\bm{r})\,.\label{eq:MNEDefinition}
\end{equation}
The equation can also be considered as the (incomplete) definitions
of the magnetizations. To make the so-defined magnetizations physically
meaningful, one needs to further require the magnetizations being
the properties of material, i.e., they should be well-behaved functions
of $\bm{r}$, and vanish outside of the sample. We also introduce
the zero-field heat magnetization: $\bm{M}_{Q}(\bm{r})\equiv\bm{M}_{E}(\bm{r})-\mu_{0}\bm{M}_{N}(\bm{r})$.

\medskip{}

\noindent \emph{Magnetizations}: We rigorously prove that, with the
appropriate current operators that follow the scaling laws Eqs.~(\ref{eq:Scaling-N}--\ref{eq:Scaling-E}),
the total magnetizations can be calculated from the following set
of equations:
\begin{align}
-\frac{\partial\bm{M}_{N}}{\partial\mu_{0}} & =\frac{\beta_{0}}{2\mathrm{i}}\left.\bm{\nabla}_{\bm{q}}\times\left\langle \hat{n}_{\bm{-\bm{q}}};\hat{\bm{J}}_{N,\bm{q}}\right\rangle _{0}\right|_{\bm{q}\rightarrow0},\label{eq:MN-mu}\\
\bm{M}_{N}-T_{0}\frac{\partial\bm{M}_{N}}{\partial T_{0}} & =\frac{\beta_{0}}{2\mathrm{i}}\left.\bm{\nabla}_{\bm{q}}\times\left\langle \hat{K}_{\bm{-\bm{q}}};\hat{\bm{J}}_{N,\bm{q}}\right\rangle _{0}\right|_{\bm{q}\rightarrow0},\label{eq:MN-T}\\
-\frac{\partial\bm{M}_{Q}}{\partial\mu_{0}} & =\frac{\beta_{0}}{2\mathrm{i}}\left.\bm{\nabla}_{\bm{q}}\times\left\langle \hat{n}_{\bm{-\bm{q}}};\hat{\bm{J}}_{Q,\bm{q}}\right\rangle _{0}\right|_{\bm{q}\rightarrow0},\label{eq:MQ-mu}\\
2\bm{M}_{Q}-T_{0}\frac{\partial\bm{M}_{Q}}{\partial T_{0}} & =\frac{\beta_{0}}{2\mathrm{i}}\left.\bm{\nabla}_{\bm{q}}\times\left\langle \hat{K}_{\bm{-\bm{q}}};\hat{\bm{J}}_{Q,\bm{q}}\right\rangle _{0}\right|_{\bm{q}\rightarrow0},\label{eq:MQ-T}
\end{align}
where $\left\langle \hat{a};\hat{b}\right\rangle _{0}\equiv(1/\beta_{0})\int_{0}^{\beta_{0}}\mathrm{d}\lambda\mathrm{Tr}\left[\hat{\rho}_{\mathrm{0}}\hat{a}(-\mathrm{i}\hbar\lambda)\hat{b}\right]$
is the Kubo canonical correlation function ($\beta_{0}\equiv1/k_{B}T_{0}$)~\cite{Kubo1983},
$\bm{M}_{N(Q)}\equiv\int\mathrm{d}\bm{r}\bm{M}_{N(Q)}(\bm{r})$, $\hat{\bm{J}}_{Q}(\bm{r})\equiv\hat{\bm{J}}_{E}(\bm{r})-\mu_{0}\hat{\bm{J}}_{N}(\bm{r})$,
and $\hat{n}_{\bm{q}}$, $\hat{K}_{\bm{q}}$, $\hat{\bm{J}}_{N,\bm{q}}$,
$\hat{\bm{J}}_{Q,\bm{q}}$ are the Fourier transform of $\hat{n}(\bm{r})$,
$\hat{K}(\bm{r)}$, $\hat{\bm{J}}_{N}(\bm{r})$, $\hat{\bm{J}}_{Q}(\bm{r})$
($\hat{a}_{\bm{q}}\equiv\int\mathrm{d}\bm{r}\hat{a}(\bm{r})e^{-\mathrm{i}\bm{q}\cdot\bm{r}}$),
respectively.

Equations (\ref{eq:MN-mu}--\ref{eq:MQ-T}) are the central results
of this Letter. The total magnetizations can be obtained by integrating
over either the chemical potential $\mu_{0}$ {[}Eqs.~(\ref{eq:MN-mu},
\ref{eq:MQ-mu}){]} or the temperature $T_{0}$ {[}Eqs.~(\ref{eq:MN-T},
\ref{eq:MQ-T}){]}. The corresponding boundary conditions are that
at $\mu_{0}\rightarrow-\infty$, $\bm{M}_{N(Q)}\rightarrow0$ and
at $T_{0}\rightarrow0$, $\bm{M}_{N}$ ($2\bm{M}_{Q}$) coincides
with right hand side (RHS) of Eq.~(\ref{eq:MN-T}){[}(\ref{eq:MQ-T}){]},
respectively. For electronic system, the two approaches are equivalent.
On the other hand, for systems without the chemical potential, such
as the phonon and magnon systems, Equation (\ref{eq:MQ-T}) is the
only option for calculating the heat (energy) magnetization. 

In Ref.~\cite{Shi2007}, a similar formula for the electromagnetic
orbital magnetization $\bm{M}\equiv-e\bm{M}_{N}$ was derived from
its thermodynamic definition $\bm{M}=-(\partial\Omega/\partial\bm{B})_{T_{0},\mu_{0}}$,
where $\Omega$ is the grand thermodynamic potential, and $\bm{B}$
is the magnetic field. It is easy to identify that RHS of Eq.~(\ref{eq:MN-T})
is nothing but $-(\partial K/\partial\bm{B})_{\mu_{0},T_{0}}$, where
$K\equiv\Omega+T_{0}S$ and $S$ is the entropy of the system. Similarly,
Eq.~(\ref{eq:MN-mu}) is just the Maxwell relation between $\partial\bm{M}/\partial\mu$
and $\partial N/\partial\bm{B}$, where $N$ is the total particle
number of the system.

One can develop a similar thermodynamic interpretation for the heat
magnetization as well. For this purpose, it is necessary to introduce
a fictitious ``magnetic field'' $\bm{B}_{s}$ which couples to $\bm{M}_{s}\equiv\bm{M}_{Q}/T_{0}$
so that $\bm{M}_{s}=-(\partial\Omega/\partial\bm{B}_{s})_{\mu_{0},T_{0}}$.
$\bm{B}_{s}$ can be related to the physical gravitomagnetic field
$\bm{B}_{g}$~\cite{Braginsky1977} by $\bm{B}_{s}\equiv-(T_{0}/c^{2})\bm{B}_{g}$.
In analogy to the particle magnetization, RHSs of Eqs.~(\ref{eq:MQ-mu}--\ref{eq:MQ-T})
are $-T_{0}(\partial N/\partial\bm{B}_{s})_{T_{0},\mu_{0}}$ and $-T_{0}(\partial K/\partial\bm{B}_{s})_{T_{0},\mu_{0}}$,
respectively, and these equations are nothing but the thermodynamic
relations. It is important to note that the particular way to introduce
the thermodynamic quantities (e.g., $\bm{M}_{s}$ instead of $\bm{M}_{Q}$)
is necessary for accounting for the extra factor of $2$ in front
of $\bm{M}_{Q}$ in Eq.~(\ref{eq:MQ-T}).

We sketch the proof of Eqs.~(\ref{eq:MN-mu}--\ref{eq:MQ-T}) in
the following~\cite{Supplemental}. We introduce the static response
functions: 
\begin{equation}
\bm{\chi}_{ij}(\bm{r},\bm{r}^{\prime})=\beta_{0}\left\langle \Delta\hat{n}_{j}(\bm{r}^{\prime});\Delta\hat{\bm{J}}_{i}(\bm{r})\right\rangle _{0},\, i,j=1,2,\label{eq:chiij definition}
\end{equation}
where $\hat{n}_{1}(\bm{r})\equiv\hat{n}(\bm{r})$, $\hat{n}_{2}(\bm{r})\equiv\hat{K}(\bm{r)}$,
$\hat{\bm{J}}_{1}(\bm{r})\equiv\hat{\bm{J}}_{N}(\bm{r})$, $\hat{\bm{J}}_{2}(\bm{r})\equiv\hat{\bm{J}}_{\bm{Q}}(\bm{r})$,
and $\Delta\hat{a}\equiv\hat{a}-\left\langle \hat{a}\right\rangle _{0}$.
Applying Eqs.~(\ref{eq:charge-continuity}--\ref{eq:energy-continuity}),
we obtain $\bm{\nabla}\cdot\bm{\chi}_{ij}(\bm{r},\bm{r}^{\prime})=(1/\mathrm{i}\hbar)\left\langle \left[\hat{n}_{j}(\bm{r}^{\prime}),\,\hat{n}_{i}(\bm{r})\right]\right\rangle _{0}$,
which implies:
\begin{equation}
\bm{\nabla}\cdot\bm{\chi}_{ij}^{\bm{q}}(\bm{r})+\mathrm{i}\bm{q}\cdot\left[\bm{\chi}_{ij}^{\bm{q}}(\bm{r})-\bm{\nabla}\times\bm{M}_{ij}(\bm{r})\right]=0\,,\label{eq:chiqij}
\end{equation}
where $\bm{\chi}_{ij}^{\bm{q}}(\bm{r})\equiv\int\mathrm{d}\bm{r}^{\prime}\bm{\chi}_{ij}(\bm{r},\bm{r}^{\prime})e^{-\mathrm{i}\bm{q}\cdot(\bm{r}-\bm{r}^{\prime})}$,
$\bm{M}_{11}(\bm{r})=0$, $\bm{M}_{12}(\bm{r})=\bm{M}_{N}(\bm{r})$,
$\bm{M}_{21}(\bm{r})=\bm{M}_{N}(\bm{r})$, $\bm{M}_{22}(\bm{r})=2\bm{M}_{Q}(\bm{r})$.
In deriving Eq.~(\ref{eq:chiqij}), we have utilized Eqs.~(\ref{eq:charge-continuity}--\ref{eq:energy-continuity})
and Eqs.~(\ref{eq:Scaling-N}--\ref{eq:Scaling-E}), which imply
the operator form of the commutators $\left[\hat{n}_{j}(\bm{r}^{\prime}),\,\hat{n}_{i}(\bm{r})\right]$.
Equation (\ref{eq:MNEDefinition}) is then used to determine the equilibrium
expectation values of the resulting commutators.

Therefore, $\bm{\chi}_{ij}^{\bm{q}}(\bm{r})$ must have the decomposition:
\begin{equation}
\bm{\chi}_{ij}^{\bm{q}}(\bm{r})=-\mathrm{i}\bm{q}\times\bm{M}_{ij}(\bm{r})+e^{-\mathrm{i}\bm{q}\cdot\bm{r}}\bm{\nabla}\times\bm{\kappa}_{ij}^{\bm{q}}(\bm{r})\,.\label{eq:chiqij1}
\end{equation}
Because both $\bm{M}_{ij}(\bm{r})$ and $\bm{\chi}_{ij}^{\bm{q}}(\bm{r})$
are properties of material, $\bm{\kappa}_{ij}^{\bm{q}}(\bm{r})$ must
also be well-behaved, i.e., it should be bounded and vanish outside
of the sample. Moreover, the value of $\bm{\kappa}_{ij}^{\bm{q}}(\bm{r})$
at the long wave limit ($\bm{q}=0$) can be related to the macroscopic
thermodynamic quantities $\left[\partial\bm{M}_{N(Q)}(\bm{r})/\partial\mu_{0}\right]_{T_{0}}$
and $\left[\partial\bm{M}_{N(Q)}(\bm{r})/\partial T_{0}\right]_{\mu_{0}}$~\cite{Supplemental}.
Applying $(\mathrm{\mathrm{i}}/2)\bm{\nabla}_{\bm{q}}\times$ to both
sides of Eq.~(\ref{eq:chiqij1}), taking limit $\bm{q}\rightarrow0$,
substituting $\bm{M}_{ij}$ and $\bm{\kappa}_{ij}^{\bm{q}=0}$ and
integrating over $\bm{r}$, we obtain Eqs.~(\ref{eq:MN-mu}--\ref{eq:MQ-T}).

\medskip{}

\noindent \emph{Thermal transport coefficients}: We can show that
the magnetizations determined by Eqs.~(\ref{eq:MN-mu}--\ref{eq:MQ-T})
will emerge naturally as corrections to the thermal transport coefficients.
To see this, we calculate the full response of the currents to small
deviation from the global equilibrium. In this case, the system can
be approximately described by the density matrix:
\begin{equation}
\hat{\rho}\approx\hat{\rho}_{\mathrm{leq}}+\hat{\rho}_{1}\,,
\end{equation}
where $\hat{\rho}_{\mathrm{leq}}$ is the local equilibrium density
matrix characterized by the local chemical potential $\mu(\bm{r})$
and local temperature $T(\bm{r})$: 
\begin{align}
\hat{\rho}_{\mathrm{leq}} & =\frac{1}{Z}\exp\left[-\int\mathrm{d}\bm{r}\frac{\hat{h}(\bm{r})-\mu(\bm{r})\hat{n}(\bm{r})}{k_{B}T(\bm{r})}\right]\,.
\end{align}
$\hat{\rho}_{1}$ is the linear response correction to the local equilibrium
density matrix, determined by the Liouville equation $\mathrm{i}\hbar\partial\hat{\rho}/\partial t+[\hat{\rho},\hat{H}_{\phi,\psi}]=0$~\cite{Mahan2000}.
We define $\alpha(\bm{r})\equiv[1+\psi(\bm{r})][\phi(\bm{r})+\mu(\bm{r})]$,
$\beta(\bm{r})\equiv1/k_{B}[1+\psi(\bm{r})]T(\bm{r})$. It is easy
to see that when $\alpha(\bm{r})$ and $\beta(\bm{r})$ are spatially
uniform, $\hat{\rho}_{\mathrm{leq}}$ becomes the exact global equilibrium
density matrix corresponding to the Hamiltonian $\hat{H}_{\phi,\psi}$,
and $\hat{\rho}_{1}=0$. Therefore, the conditions of the global equilibrium
are $\bm{\nabla}\alpha(\bm{r})=0$ and $\bm{\nabla}\beta(\bm{r})=0$~\cite{Luttinger1964}.

We define $\bm{J}_{1}^{\phi,\psi}=\bm{J}_{N}^{\phi,\psi}$ and $\bm{J}_{2}^{\phi,\psi}=\hat{\bm{J}}_{Q}^{\phi,\psi}\equiv\bm{J}_{E}^{\phi,\psi}-\alpha(\bm{r})\bm{J}_{N}^{\phi,\psi}$.
The forces conjugate to these currents are $\bm{X}_{1}=-\beta(\bm{r})\bm{\nabla}\alpha(\bm{r})$
and $\bm{X}_{2}=\bm{\nabla}\beta(\bm{r})$, respectively, so that
the entropy generation is $\partial s/\partial t+\bm{\nabla}\cdot(\beta\bm{J}_{Q}^{\phi,\psi})=\sum_{i}\bm{J}_{i}^{\phi,\psi}\cdot\bm{X}_{i}$~\cite{deGroot1984}.
The expectation values of the currents have two parts of contributions:
\begin{align}
\bm{J}_{i}^{\phi,\psi} & =\bm{J}_{i}^{\mathrm{leq}}+\bm{J}_{i}^{\mathrm{Kubo}}\,,\label{eq:expectation value}
\end{align}
where $\bm{J}_{i}^{\mathrm{Kubo}}\equiv\mathrm{Tr}\hat{\rho}_{\mathrm{1}}\hat{\bm{J}}_{i}^{\phi,\psi}$
is just the usual linear response contribution with the response coefficients
determinable by the Kubo formula~\cite{Mahan2000}. Besides this,
there is an extra contribution $\bm{J}_{i}^{\mathrm{leq}}=\mathrm{Tr}\hat{\rho}_{\mathrm{leq}}\hat{\bm{J}}_{i}^{\phi,\psi}\,,$
which is due to the inhomogeneous local chemical potential and temperature
field. We assume that the deviation from the homogeneity is small
so that $\mu(\bm{r})\approx\mu_{0}+\delta\mu(\bm{r})$, $1/T(\bm{r})\approx(1/T_{0})+\delta[1/T(\bm{r})]$.
By applying the static response theory~\cite{Kubo1983}, we obtain,
to the linear order of $x_{1}(\bm{r})\equiv\delta\mu(\bm{r})$ and
$x_{2}(\bm{r})\equiv-T_{0}\delta[1/T(\bm{r})]$:
\begin{multline}
\bm{J}_{i}^{\mathrm{leq}}(\bm{r})\approx\bm{J}_{i}^{\mathrm{eq}}(\bm{r})+\sum_{j=1}^{2}\int\mathrm{d}\bm{r}^{\prime}\bm{\chi}_{ij}(\bm{r},\bm{r}^{\prime})x_{j}(\bm{r}^{\prime})\,,\label{eq:jleq}
\end{multline}
where $\bm{\chi}_{ij}(\bm{r},\bm{r}^{\prime})$ is the static response
function defined in Eq.~(\ref{eq:chiij definition}), and $\bm{J}_{i}^{\mathrm{eq}}(\bm{r})\equiv\left\langle \hat{\bm{J}}_{i}^{\phi,\psi}(\bm{r})\right\rangle _{0}$,
which can be determined by Eq.~(\ref{eq:MNEDefinition}) and Eqs.~(\ref{eq:Scaling-N}--\ref{eq:Scaling-E}).
Substituting Eq.~(\ref{eq:chiqij1}) into Eq.~(\ref{eq:jleq}),
and after some algebra, we obtain, to the linear order of $\phi$,
$\psi$, $\delta\mu$ and $\delta(1/T)$~\cite{Supplemental},
\begin{align}
\bm{J}_{1}^{\mathrm{leq}}(\bm{r})\approx & \bm{\nabla}\times\bm{M}_{N}^{\phi,\psi}(\bm{r})-\frac{1}{\beta}\bm{M}_{N}(\bm{r})\times\bm{X}_{2}\,,\label{eq:J1leqX}\\
\bm{J}_{2}^{\mathrm{leq}}(\bm{r})\approx & \bm{\nabla}\times\bm{M}_{E}^{\phi,\psi}(\bm{r})-\alpha(\bm{r})\bm{\nabla}\times\bm{M}_{N}^{\phi,\psi}(\bm{r})\nonumber \\
 & -\frac{1}{\beta}\bm{M}_{N}(\bm{r})\times\bm{X}_{1}-\frac{2}{\beta}\bm{M}_{Q}(\bm{r})\times\bm{X}_{2}\,,\label{eq:J2leqX}
\end{align}
where $\bm{M}_{N}^{\phi,\psi}(\bm{r})\equiv\left[1+\psi(\bm{r})\right]\bm{M}_{N}(\bm{r})+\delta\bm{M}_{N}(\bm{r})$,
$\bm{M}_{E}^{\phi,\psi}(\bm{r})\equiv\left[1+\psi(\bm{r})\right]^{2}[\bm{M}_{E}(\bm{r})+\phi(\bm{r})\bm{M}_{N}(\bm{r})]+\delta\bm{M}_{E}(\bm{r})$,
and $\delta\bm{M}_{N(E)}(\bm{r})$ is the correction to the particle
(energy) magnetization due to the spatial gradients of the chemical
potential and temperature, determinable by $\kappa_{ij}^{\bm{q}}(\bm{r})$. 

Applying Eqs.~(\ref{eq:expectation value}, \ref{eq:J1leqX}, \ref{eq:J2leqX}),
we can obtain the total currents responding to the non-equilibrium
forces. However, due to the presence of $\bm{J}_{i}^{\mathrm{leq}}$,
such responses break the fundamental non-equilibrium thermodynamic
relations~\cite{deGroot1984}: (1) Onsager reciprocal relations;
(2) Einstein relations, i. e., the currents should only be proportional
to $\bm{\nabla}\alpha$ and $\bm{\nabla}\beta$, and vanish when the
system is in the global equilibrium. The problem can be remedied by
defining the transport currents as $\bm{J}_{N(E)}^{\phi,\psi,\mathrm{tr}}=\bm{J}_{N(E)}^{\phi,\psi}-\bm{\nabla}\times\bm{M}_{N(E)}^{\phi,\psi}$,
and the corresponding transport responses then become:
\begin{equation}
\left[\begin{array}{c}
\bm{J}_{1}^{\mathrm{tr}}\\
\bm{J}_{2}^{\mathrm{tr}}
\end{array}\right]=\left[\begin{array}{cc}
\overleftrightarrow{L}^{(11)} & \overleftrightarrow{L}^{(12)}-\frac{\bm{M}_{N}}{\beta_{0}V}\times\\
\overleftrightarrow{L}^{(21)}-\frac{\bm{M}_{N}}{\beta_{0}V}\times & \overleftrightarrow{L}^{(22)}-\frac{2\bm{M}_{Q}}{\beta_{0}V}\times
\end{array}\right]\left[\begin{array}{c}
\bm{X}_{1}\\
\bm{X}_{2}
\end{array}\right],\label{eq:Linear response}
\end{equation}
where $\bm{J}_{i}^{\mathrm{tr}}\equiv(1/V)\int\mathrm{d}\bm{r}\bm{J}_{i,}^{\phi,\psi,\mathrm{tr}}(\bm{r})$,
and $V$ is the total volume of the system. $\overleftrightarrow{L}^{\left(ij\right)}$
is a tensor of rank two with the component $L_{\alpha\gamma}^{(ij)}=\int_{0}^{\infty}\mathrm{d}te^{-st}\left\langle \hat{J}_{j,\gamma};\hat{J}_{i,\alpha}(t)\right\rangle _{0}$
($\alpha,\gamma=x,y,z$), which is the usual response coefficient
determined by the Kubo formula~\cite{Mahan2000}. It is easy to verify
that both the Onsager relations and the Einstein relations are recovered.
The magnetizations determined in Eqs.~(\ref{eq:MN-mu}--\ref{eq:MQ-T})
naturally emerge as the corrections to the thermal transport coefficients.

\medskip{}

\noindent \emph{Application}: We can apply these general results to
study the thermal Hall coefficient of a non-interacting anomalous
Hall system~\cite{Jungwirth2002,Xiao2006}, and show how the unphysical
divergence is eliminated and the Wiedemann-Franz law emerges. The
energy density of such a system, in the presence of the external fields
$\phi(\bm{r})$ and $\psi(\bm{r})$, can in general be written as:
\begin{multline}
\hat{h}^{\phi,\psi}(\bm{r})=\left[1+\psi(\bm{r})\right]\left\{ \frac{m}{2}\left[\hat{\bm{v}}\hat{\varphi}(\bm{r})\right]^{\dagger}\cdot\left[\hat{\bm{v}}\hat{\varphi}(\bm{r})\right]\right.\\
\left.+\hat{\varphi}^{\dagger}(\bm{r})\left[V(\bm{r})+\phi(\bm{r})\right]\hat{\varphi}(\bm{r})\right\} \,,
\end{multline}
where $\hat{\varphi}(\bm{r})$ ($\hat{\varphi}^{\dagger}(\bm{r})$)
is the electron annihilation (creation) field operator with the two
spin components, $ $$\hat{\bm{v}}\equiv(1/m)[-\mathrm{i}\hbar\bm{\nabla}+\bm{A}_{\mathrm{so}}(\bm{r})]$
is the velocity operator with $\bm{A}_{\mathrm{so}}(\bm{r})$ being
the non-abelian gauge potential characterizing the spin-orbit coupling,
and $V(\bm{r})$ is the periodic potential. The field operator $\hat{\varphi}(\bm{r})$
satisfies the Schr\"{o}dinger equation: $\mathrm{i}\hbar\partial\hat{\varphi}/\partial t=\hat{\mathcal{H}}_{\phi,\psi}\hat{\varphi}$
with $\hat{\mathcal{H}}_{\phi,\psi}=(m/2)\hat{\bm{v}}\cdot\left[1+\psi(\bm{r})\right]\hat{\bm{v}}+\left[1+\psi(\bm{r})\right]\left[V(\bm{r})+\phi(\bm{r})\right]$.
An appropriate energy current operator that does satisfy both Eq.~(\ref{eq:energy-continuity})
and the scaling law Eq.~(\ref{eq:Scaling-E}) is, 
\begin{multline}
\hat{\bm{J}}_{E}^{\phi,\psi}(\bm{r})=\frac{1+\psi}{2}\left[\left(\hat{\bm{v}}\hat{\varphi}\right)^{\dagger}\left(\hat{\mathcal{H}}_{\phi,\psi}\hat{\varphi}\right)+\left(\hat{\mathcal{H}}_{\phi,\psi}\hat{\varphi}\right)^{\dagger}\left(\hat{\bm{v}}\hat{\varphi}\right)\right]\\
+\frac{\mathrm{i}\hbar}{8}\bm{\nabla}\times\left[\left(1+\psi\right)^{2}\left(\hat{\bm{v}}\hat{\varphi}\right)^{\dagger}\times\left(\hat{\bm{v}}\hat{\varphi}\right)\right]\,.\label{eq:JEproper}
\end{multline}
The presence of the last term is essential for satisfying the scaling
law Eq.~(\ref{eq:Scaling-E}).

With the appropriate energy current operator at hand, we calculate
the thermal Hall coefficient. The usual Kubo formula yields,
\begin{equation}
\kappa_{xy}^{\mathrm{Kubo}}\equiv\frac{L_{xy}^{(22)}}{k_{B}T_{0}^{2}}=\frac{1}{2T_{0}\hbar V}\sum_{n\bm{k}}\Pi_{n\bm{k}}^{z}f_{n\bm{k}},\label{eq:kappaKubo}
\end{equation}
where $\Pi_{n\bm{k}}^{z}=\mathrm{Im}\left\langle \frac{\partial u_{n\bm{k}}}{\partial k_{x}}\right|\left(\hat{\mathcal{H}}{}_{\bm{k}}+\epsilon_{n\bm{k}}-2\mu_{0}\right)^{2}\left|\frac{\partial u_{n\bm{k}}}{\partial k_{y}}\right\rangle $,
$u_{n\bm{k}}$ is the periodic part of Bloch wave function for band
$n$ and quasi-momentum $\bm{k}$, $f_{n\bm{k}}\equiv f(\epsilon_{n\bm{k}})$
is the Fermi distribution function, $\hat{\mathcal{H}}_{\bm{k}}=(1/2m)[-\mathrm{i}\hbar\bm{\nabla}+\bm{A}_{\mathrm{so}}(\bm{r})+\hbar\bm{k}]^{2}+V(\bm{r})$,
and $\epsilon_{n\bm{k}}$ is the electron dispersion~\cite{Katsura2010}.
It is easy to see that the coefficient diverges at the zero temperature.

We calculate $\tilde{M}_{Q}^{z}\equiv(\beta_{0}/2\mathrm{i})\left.\bm{\nabla}_{\bm{q}}\times\left\langle \hat{K}_{\bm{-\bm{q}}};\hat{\bm{J}}_{Q,\bm{q}}\right\rangle _{0}\right|_{z,\bm{q}\rightarrow0}$,
and obtain:
\begin{multline}
\tilde{M}_{Q}^{z}=-\frac{1}{4\hbar}\sum_{n\bm{k}}\biggl\{\Pi_{n\bm{k}}^{z}\left[2f_{n\bm{k}}+\left(\epsilon_{n\bm{k}}-\mu\right)f_{n\bm{k}}^{\prime}\right]\\
+2\Omega_{n\bm{k}}^{z}\left(\epsilon_{n\bm{k}}-\mu_{0}\right)^{3}f_{n\bm{k}}^{\prime}\biggr\}\,,
\end{multline}
where $\Omega_{n\bm{k}}^{z}\equiv-2\mathrm{Im}\left\langle \partial u_{n\bm{k}}/\partial k_{x}\left|\partial u_{n\bm{k}}/\partial k_{y}\right.\right\rangle $.
$M_{Q}^{z}$ is obtained by integrating Eq.~(\ref{eq:MQ-T}). After
some algebra, we obtain $\kappa_{xy}^{\mathrm{tr}}\equiv\kappa_{xy}^{\mathrm{Kubo}}+(2M_{Q}^{z}/T_{0}V)$:
\begin{equation}
\kappa_{xy}^{\mathrm{tr}}=-\frac{1}{e^{2}T_{0}}\int\mathrm{d}\epsilon(\epsilon-\mu_{0})^{2}\sigma_{xy}(\epsilon)f^{\prime}(\epsilon)\,,
\end{equation}
where $\sigma_{xy}(\epsilon)=-(e^{2}/\hbar)\sum_{\epsilon_{n\bm{k}}\leq\epsilon}\Omega_{n\bm{k}}^{z}$
is the zero temperature anomalous Hall coefficient for a system with
the chemical potential $\epsilon$~\cite{Jungwirth2002,Xiao2006}.
It recovers the Wiedemann-Franz law at the low temperature $k_{B}T_{0}\ll\mu_{0}$~\cite{Ashcroft1976,Smrcka1977},
and the unphysical divergence is eliminated.\medskip{}

In summary, we have developed a systematic approach for calculating
the particle and heat (energy) magnetizations. We also explicitly
show that these magnetizations naturally emerges as the corrections
to the thermal transport coefficients, recovering the Onsager and
Einstein relations, and eliminating the unphysical divergences. Our
approach make no assumption on the nature of the system, so is equally
applicable to fermionic (e.g., electron) or bosonic (e.g., phonon,
magnon) systems, either non-interacting or interacting. The approach
does not involve the ill-defined spatially extended operators, so
is usable in practical calculations.


\begin{thebibliography}{References}
\bibitem{deGroot1984}S. R. de Groot and P. Mazur, \emph{Non-equilibrium
Thermodynamics,} (Dover Publications, 1984).

\bibitem{Strohm2005}C. Strohm, G. L. J. A. Rikken, and P. Wyder,
Phys. Rev. Lett. \textbf{95}, 155901 (2005).

\bibitem{Inyushkin2007}A.V. Inyushkin and A. N. Taldenkov, JETP Lett.
\textbf{86}, 379 (2007).

\bibitem{Onose2010}Y. Onose, T. Ideue, H. Katsura, Y. Shiomi, N.
Nagaosa, and Y. Tokura, Science \textbf{329}, 297 (2010).

\bibitem{Sheng2006}L. Sheng, D. N. Sheng, and C. S. Ting, Phys. Rev.
Lett. \textbf{96}, 155901(2006)

\bibitem{Kagan2008}Yu. Kagan and L. A. Maksimov, Phys. Rev. Lett.
\textbf{100}, 145902 (2008)

\bibitem{Zhang2010}Lifa Zhang, Jie Ren, Jian-Sheng Wang and Baowen
Li, Phys. Rev. Lett. \textbf{105}, 225901(2010)

\bibitem{Katsura2010}H. Katsura, N. Nagaosa, and P. A. Lee, Phys.
Rev. Lett. \textbf{104}, 066403 (2010). 

\bibitem{Agarwalla2011}B.K. Agarwalla, Lifa Zhang, Jian-Sheng Wang
and Baowen Li, Eur. Phys. J. B \textbf{81}, 197 (2011)

\bibitem{Matsumoto2011}R. Matsumoto and S. Murakami, Phys. Rev. Lett.
\textbf{106}, 197202 (2011); arXiv:1106.1987 (2011).

\bibitem{Smrcka1977}L. Smr\v{c}ka, and P. St\v{r}eda, J. Phys. C
\textbf{10}, 2153 (1977).

\bibitem{Cooper1997}N.R. Cooper, B.I. Halperin, and I.M. Ruzin, Phys.
Rev. B \textbf{55}, 2344 (1997). 

\bibitem{Ryu2010}S. Ryu, J.E. Moore, and A.W.W. Ludwig, arXiv:1010.0936
(2010). 

\bibitem{Hirst1997}L.L. Hirst, Rev. Mod. Phys. \textbf{69}, 607 (1997).

\bibitem{Xiao2006}D. Xiao, Y. Yao, Z. Fang, and Q. Niu, Phys. Rev.
Lett. \textbf{97}, 026603 (2006). 

\bibitem{Onose2008}Y. Onose, Y. Shiomi, and Y. Tokura, Phys. Rev.
Lett. \textbf{100}, 016601 (2008). 

\bibitem{Luttinger1964}J. M. Luttinger, Phys. Rev. \textbf{135},
A1505 (1964). 

\bibitem{Kubo1983}R. Kubo, M. Toda and N. Hashitsume, \emph{Statistical
Physics II}, (Springer-Verlag, 1983).

\bibitem{Shi2007}J.R. Shi, G. Vignale, D. Xiao, and Q. Niu, Phys.
Rev. Lett. \textbf{99}, 197202 (2007). 

\bibitem{Braginsky1977}V. B. Braginsky, C. M. Caves, and K. S. Thorne,
Phys. Rev. D \textbf{15}, 2047 (1977). 

\bibitem{Supplemental}The details of derivations are presented in
the supplemental text.

\bibitem{Mahan2000}G.D. Mahan, \emph{Many-Particle Physics, Third
Edition}, (Kluwer Academic, 2000).

\bibitem{Jungwirth2002}T. Jungwirth, Q. Niu, and A.H. MacDonald,
Phys. Rev. Lett. \textbf{88}, 207208 (2002). 

\bibitem{Ashcroft1976}N.W. Ashcroft and N.D. Mermin, \emph{Solid
State Physics}, (Harcourt College, 1976)\end{thebibliography}
\end{document}


\title{Supplementary information for ``Energy Magnetization and Thermal
Hall Effect''}

\author{Tao Qin}

\affiliation{Institute of Physics, Chinese Academy of Sciences, Beijing 100190,
China}

\author{Qian Niu }

\affiliation{Department of Physics, The University of Texas at Austin, Austin,
Texas 78712, USA}

\affiliation{International Center for Quantum Materials, Peking University, Beijing
100871, China}

\author{Junren Shi}

\affiliation{International Center for Quantum Materials, Peking University, Beijing
100871, China}

\maketitle
\tableofcontents{}

\section{Kubo's Canonical Correlation Function}

Kubo's canonical correlation function is defined as~\cite{Kubo1983}:
\begin{equation}
\left\langle \hat{B};\hat{A}\right\rangle _{0}=\frac{1}{\beta_{0}}\mathrm{Tr}\left[\hat{\rho}_{\mathrm{eq}}\int_{0}^{\beta_{0}}d\lambda e^{\lambda\hat{\mathscr{H}}}\hat{B}e^{-\lambda\hat{\mathscr{H}}}\hat{A}\right]\,,
\end{equation}
where $\hat{\mathscr{H}}$ is the Hamiltonian of the system, and $\hat{\rho}_{\mathrm{eq}}=(1/Z)\exp\left(-\beta_{0}\hat{\mathscr{H}}\right)$.
Some of its properties used in the main text are~\cite{Kubo1983}:
\begin{align}
\left\langle \Delta\hat{A};\Delta\hat{B}\right\rangle _{0} & =\left\langle \Delta\hat{B};\Delta\hat{A}\right\rangle _{0}\,,
\end{align}
and: 
\begin{align}
\beta_{0}\left\langle \Delta\hat{\dot{A}};\Delta\hat{B}\right\rangle _{0}= & -\beta_{0}\left\langle \Delta\hat{A};\Delta\hat{\dot{B}}\right\rangle _{0}=\frac{1}{\mathrm{i}\hbar}\left\langle \left[\Delta\hat{A},\,\Delta\hat{B}\right]\right\rangle _{0}\,,\label{eq:correlation2commutator}
\end{align}
where $\Delta\hat{A}\equiv\hat{A}-\left\langle \hat{A}\right\rangle _{0}$
and $\hat{\dot{A}}\equiv(1/\mathrm{i}\hbar)[\hat{A},\hat{\mathscr{H}}]$.

For a system perturbed by a static external force:
\begin{equation}
\hat{\mathscr{H}}^{\prime}=\hat{\mathscr{H}}-\hat{A}\delta x\,.
\end{equation}
The change of the expectation value of an operator $\hat{B}$ can
be calculated to the linear order~\cite{Kubo1983}:
\begin{equation}
\delta\left\langle \hat{B}\right\rangle =\left\langle \hat{B}\right\rangle _{\delta x}-\left\langle \hat{B}\right\rangle _{0}\equiv\chi_{BA}\delta x\,,\label{eq:chiBA}
\end{equation}
where: 
\begin{equation}
\chi_{BA}\equiv\beta_{0}\left\langle \Delta\hat{B};\Delta\hat{A}\right\rangle _{0}\,.
\end{equation}

\section{Details of Derivation for Eqs.~(7-10), Magnetization Formula}

The derivation of Eqs.~(7-10) is detailed in the following:

(1) We introduce $\bm{\chi}_{ij}\left(\bm{r},\bm{r}^{\prime}\right)=\beta_{0}\left\langle \Delta\hat{n}_{j}\left(\bm{r}^{\prime}\right);\,\Delta\hat{\bm{J}}_{i}\left(\bm{r}\right)\right\rangle _{0}$
with $i,j=1,2$, where $\hat{n}_{1}(\bm{r})\equiv\hat{n}(\bm{r})$,
$\hat{n}_{2}(\bm{r})\equiv\hat{K}(\bm{r)}$, $\hat{\bm{J}}_{1}(\bm{r})\equiv\hat{\bm{J}}_{N}(\bm{r})$,
$\hat{\bm{J}}_{2}(\bm{r})\equiv\hat{\bm{J}}_{Q}(\bm{r})$, $\Delta\hat{a}\equiv\hat{a}-\left\langle \hat{a}\right\rangle _{0}$
and $\left\langle \cdots\right\rangle _{0}\equiv\mathrm{Tr}\left[\hat{\rho}_{0}\cdots\right]$,
so we have:
\begin{align}
\bm{\nabla}\cdot\bm{\chi}_{ij}\left(\bm{r},\bm{r}^{\prime}\right)= & \beta_{0}\left\langle \Delta\hat{n}_{j}\left(\bm{r}^{\prime}\right);\,\bm{\nabla}\cdot\Delta\hat{\bm{J}}_{i}\left(\bm{r}\right)\right\rangle _{0}\,,\label{eq:chi1}\\
= & -\beta_{0}\left\langle \hat{n}_{j}\left(\bm{r}^{\prime}\right);\,\hat{\dot{n}}_{i}\left(\bm{r}\right)\right\rangle _{0}\,.\label{eq:chi2}
\end{align}
From Eq.~(\ref{eq:chi1}) to Eq.~(\ref{eq:chi2}), we have used
$\bm{\nabla}\cdot\hat{\bm{J}}_{i}\left(\bm{r}\right)=-\hat{\dot{n}}_{i}$
and $\bm{\nabla}\cdot\bm{J}_{i}^{\mathrm{eq}}=0$. Using Eq.~(\ref{eq:correlation2commutator}),
we obtain: 
\begin{equation}
\bm{\nabla}\cdot\bm{\chi}_{ij}\left(\bm{r},\bm{r}^{\prime}\right)=\frac{1}{\mathrm{i}\hbar}\left\langle \left[\hat{n}_{j}\left(\bm{r}^{\prime}\right),\,\hat{n}_{i}\left(\bm{r}\right)\right]\right\rangle _{0}\,.\label{eq:nablachi}
\end{equation}
We define $\bm{\chi}_{ij}^{\bm{q}}\left(\bm{r}\right)\equiv\int d\bm{r}^{\prime}\bm{\chi}_{ij}\left(\bm{r},\bm{r}^{\prime}\right)e^{-\mathrm{i}\bm{q}\cdot\left(\bm{r}-\bm{r}^{\prime}\right)}$,
and have: 
\begin{equation}
\bm{\nabla}\cdot\bm{\chi}_{ij}^{\bm{q}}\left(\bm{r}\right)+\mathrm{i}\bm{q}\cdot\bm{\chi}_{ij}^{\bm{q}}\left(\bm{r}\right)=\frac{1}{\mathrm{i}\hbar}\int d\bm{r}^{\prime}\left\langle \left[\hat{n}_{j}\left(\bm{r}^{\prime}\right),\,\hat{n}_{i}\left(\bm{r}\right)\right]\right\rangle _{0}e^{-\mathrm{i}\bm{q}\cdot\left(\bm{r}-\bm{r}^{\prime}\right)}\,,\label{eq:commutator}
\end{equation}
where $i,j=1,2$. 

(2) We can obtain the right hand side of Eq.~(\ref{eq:commutator})
from the definitions of currents and their scaling laws. Basically,
we have:
\begin{align}
\frac{1}{\mathrm{i}\hbar}[\hat{n}(\bm{r}),\,\hat{H}^{\phi,\psi}]= & -\bm{\nabla}\cdot\hat{\bm{J}}_{N}^{\phi,\psi}(\bm{r})\,,\label{eq:particle}\\
\frac{1}{\mathrm{i}\hbar}[\hat{h}^{\phi,\psi}(\bm{r}),\,\hat{H}^{\phi,\psi}]= & -\bm{\nabla}\cdot\hat{\bm{J}}_{E}^{\phi,\psi}(\bm{r})\,,\label{eq:energy}
\end{align}
and:
\begin{align}
\hat{\bm{J}}_{N}^{\phi,\psi}(\bm{r}) & =\left[1+\psi(\bm{r})\right]\hat{\bm{J}}_{N}(\bm{r})\,,\label{eq:scaleN}\\
\hat{\bm{J}}_{E}^{\phi,\psi}(\bm{r}) & =\left[1+\psi(\bm{r})\right]^{2}\left[\hat{\bm{J}}_{E}(\bm{r})+\phi\left(\bm{r}\right)\hat{\bm{J}}_{N}(\bm{r})\right]\,.\label{eq:scaleE}
\end{align}

If we set $\phi\left(\bm{r}\right)=0$ and $1+\psi\left(\bm{r}\right)=e^{\mathrm{i}\bm{q}\cdot\bm{r}}$,
Eq.~(\ref{eq:particle}) becomes: 
\begin{align}
\frac{1}{\mathrm{i}\hbar}\int d\bm{r}^{\prime}\left[e^{\mathrm{i}\bm{q}\cdot\bm{r}^{\prime}}\hat{h}\left(\bm{r}^{\prime}\right),\,\hat{n}\left(\bm{r}\right)\right] & =\bm{\nabla}\cdot\left(e^{\mathrm{i}\bm{q}\cdot\bm{r}}\hat{\bm{J}}_{N}\left(\bm{r}\right)\right)\,,\\
 & =e^{\mathrm{i}\bm{q}\cdot\bm{r}}\mathrm{i}\bm{q}\cdot\hat{\bm{J}}_{N}\left(\bm{r}\right)+e^{\mathrm{i}\bm{q}\cdot\bm{r}}\bm{\nabla}\cdot\hat{\bm{J}}_{N}\left(\bm{r}\right)\,,
\end{align}
and we obtain:
\begin{equation}
\frac{1}{\mathrm{i}\hbar}\int d\bm{r}^{\prime}\left\langle \left[e^{\mathrm{i}\bm{q}\cdot\left(\bm{r}^{\prime}-\bm{r}\right)}\hat{h}\left(\bm{r}^{\prime}\right),\,\hat{n}\left(\bm{r}\right)\right]\right\rangle _{0}=\mathrm{i}\bm{q}\cdot\bm{\nabla}\times\bm{M}_{N}\left(\bm{r}\right)\,,
\end{equation}
where we have used $\bm{\nabla}\cdot\hat{\bm{J}}_{N}^{\mathrm{eq}}\left(\bm{r}\right)=0$
and $\hat{\bm{J}}_{N}^{\mathrm{eq}}\left(\bm{r}\right)=\bm{\nabla}\times\bm{M}_{N}\left(\bm{r}\right)$.
This is exactly the right hand side of Eq.~(\ref{eq:commutator})
for $i=1$, $j=2$.

Using the similar approach, we can prove that:
\begin{equation}
\frac{1}{\mathrm{i}\hbar}\int d\bm{r}^{\prime}\left\langle \left[\hat{n}_{j}\left(\bm{r}^{\prime}\right),\,\hat{n}_{i}\left(\bm{r}\right)\right]\right\rangle _{0}e^{-\mathrm{i}\bm{q}\cdot\left(\bm{r}-\bm{r}^{\prime}\right)}=\mathrm{i}\bm{q}\cdot\bm{\nabla}\times\bm{M}_{ij}\left(\bm{r}\right)\,,
\end{equation}
where $\bm{M}_{11}(\bm{r})=0$, $\bm{M}_{12}(\bm{r})=\bm{M}_{N}(\bm{r})$,
$\bm{M}_{21}(\bm{r})=\bm{M}_{N}(\bm{r})$, and $\bm{M}_{22}(\bm{r})=2\bm{M}_{Q}(\bm{r})$.

(3) Therefore, $\chi_{ij}^{\bm{q}}(\bm{r})$ satisfies the equation:
\begin{equation}
\bm{\nabla}\cdot\bm{\chi}_{ij}^{\bm{q}}\left(\bm{r}\right)+\mathrm{i}\bm{q}\cdot\left[\bm{\chi}_{ij}^{\bm{q}}\left(\bm{r}\right)-\bm{\nabla}\times\bm{M}_{ij}\left(\bm{r}\right)\right]=0\,,\label{eq:nablachiq}
\end{equation}
 and it has the general solution: 
\begin{equation}
\bm{\chi}_{ij}^{\bm{q}}\left(\bm{r}\right)=-\mathrm{i}\bm{q}\times\bm{M}_{ij}\left(\bm{r}\right)+e^{-\mathrm{i}\bm{q}\cdot\bm{r}}\bm{\nabla}\times\bm{\kappa}_{ij}^{\bm{q}}\left(\bm{r}\right)\,,\label{eq:chiq}
\end{equation}
where $\bm{\kappa}_{ij}^{\bm{q}}(\bm{r})$ is an arbitrary function.
This equation can be considered as a decomposition of $\chi_{ij}^{\bm{q}}(\bm{r})$.
It is important to note that the decomposition is not necessary to
be unique, because the magnetization can only be defined up to a gradient.
However, the arbitrariness does not affect our result on the total
magnetizations, as long as both $\bm{M}_{ij}(\bm{r})$ and $\bm{\kappa}_{ij}^{\bm{q}}\left(\bm{r}\right)$
are well behaved functions: i.e., they are bounded for all $\bm{r}$.
The constraint is necessary because, first, magnetizations are properties
of materials; second, only when these functions are well behaved,
can their contributions presented in Eq.~(18-19) be well defined.

(4) We can relate $\bm{\kappa}_{ij}^{\bm{q}=0}\left(\bm{r}\right)$
to the macroscopic thermodynamic quantities. To see this, we use Eq.~(17)
and see how the equilibrium currents are perturbed by the spatially
uniform changes of the chemical potential and the temperature. We
have:

\begin{align}
\delta\bm{J}_{i}^{\mathrm{eq}}(\bm{r}) & \approx\int\mathrm{d}\bm{r}^{\prime}\left[\bm{\chi}_{i1}(\bm{r},\bm{r}^{\prime})\delta\mu_{0}-\bm{\chi}_{i2}(\bm{r},\bm{r}^{\prime})T_{0}\delta(1/T_{0})\right]\,,\\
 & =\bm{\chi}_{i1}^{\bm{q}=0}(\bm{r})\delta\mu_{0}-\bm{\chi}_{i2}^{\bm{q}=0}(\bm{r})T_{0}\delta(1/T_{0})\,,\\
 & =\bm{\nabla}\times\left[\bm{\kappa}_{i1}^{\bm{q}=0}(\bm{r})\delta\mu_{0}-\bm{\kappa}_{i2}^{\bm{q}=0}(\bm{r})T_{0}\delta(1/T_{0})\right]\,.
\end{align}
Note that $\delta\bm{J}_{2}^{\mathrm{eq}}(\bm{r})\equiv\mathrm{Tr}\left[\bm{J}_{E}^{\mathrm{\phi,\psi}}(\bm{r})\delta\hat{\rho}_{\mathrm{leq}}\right]-\alpha(\bm{r})\left[\bm{J}_{N}^{\mathrm{\phi,\psi}}(\bm{r})\delta\hat{\rho}_{\mathrm{leq}}\right]\approx\delta\bm{J}_{E}^{\mathrm{eq}}(\bm{r})-\mu_{0}\delta\bm{J}_{N}^{\mathrm{eq}}(\bm{r})$.
On the other hand, $\delta\bm{J}_{i}^{\mathrm{eq}}$ is, by definition:
\begin{align}
\delta\bm{J}_{1}^{\mathrm{eq}} & =\bm{\nabla}\times\left(\delta\bm{M}_{N}\right)\,,\\
\delta\bm{J}_{2}^{\mathrm{eq}} & =\bm{\nabla}\times\left(\delta\bm{M}_{E}\right)-\mu_{0}\bm{\nabla}\times\left(\delta\bm{M}_{N}\right)\,.
\end{align}
Comparing the two sides, we obtain:
\begin{align}
\bm{\kappa}_{11}^{\bm{q}=0}\left(\bm{r}\right)= & \left.\frac{\partial\bm{M}_{N}\left(\bm{r}\right)}{\partial\mu_{0}}\right|_{T_{0}}\,,\\
\bm{\kappa}_{12}^{\bm{q}=0}\left(\bm{r}\right)= & T_{0}\left.\frac{\partial\bm{M}_{N}\left(\bm{r}\right)}{\partial T_{0}}\right|_{\mu_{0}}\,,\\
\bm{\kappa}_{21}^{\bm{q}=0}\left(\bm{r}\right)= & \left.\frac{\partial\bm{M}_{Q}\left(\bm{r}\right)}{\partial\mu_{0}}\right|_{T_{0}}+\bm{M}_{N}\left(\bm{r}\right)\,,\\
\bm{\kappa}_{22}^{\bm{q}=0}\left(\bm{r}\right)= & T_{0}\left.\frac{\partial\bm{M}_{Q}\left(\bm{r}\right)}{\partial T_{0}}\right|_{\mu_{0}}\,.
\end{align}

(5) Equation (\ref{eq:chiq}) can be rewritten as:
\begin{equation}
\bm{\chi}_{ij}^{\bm{q}}\left(\bm{r}\right)=-\mathrm{i}\bm{q}\times\left[\bm{M}_{ij}\left(\bm{r}\right)-e^{-\mathrm{i}\bm{q}\cdot\bm{r}}\bm{\kappa}_{ij}^{\bm{q}}\left(\bm{r}\right)\right]+\bm{\nabla}\times\left[e^{-\mathrm{i}\bm{q}\cdot\bm{r}}\bm{\kappa}_{ij}^{\bm{q}}\left(\bm{r}\right)\right]\,,\label{eq:chiq-2}
\end{equation}
Applying $\bm{\nabla}_{\bm{q}}\times$ to the both sides of Eq.~(\ref{eq:chiq-2})
and setting $\bm{q}\rightarrow0$, we obtain: 
\begin{align}
\frac{\mathrm{i}}{2}\left.\bm{\nabla}_{\bm{q}}\times\bm{\chi}_{ij}^{\bm{q}}\left(\bm{r}\right)\right|_{\bm{q}\rightarrow0} & =-\bm{M}_{ij}\left(\bm{r}\right)+\bm{\kappa}_{ij}^{\bm{q}=0}\left(\bm{r}\right)-\bm{\nabla}\times\bm{U}_{ij}\left(\bm{r}\right)\,,\label{eq:nablaqchiij}
\end{align}
where $\bm{U}_{ij}\left(\bm{r}\right)=\left.\frac{\mathrm{i}}{2}\bm{\nabla}_{\bm{q}}\times\left(e^{-\mathrm{i}\bm{q}\cdot\bm{r}}\bm{\kappa}_{ij}^{\bm{q}}\left(\bm{r}\right)\right)\right|_{\bm{q}\rightarrow0}$.
After substituting different components of $\bm{M}_{ij}\left(\bm{r}\right)$
and $\bm{\kappa}_{ij}^{\bm{q}=0}\left(\bm{r}\right)$ and integrating
over $\bm{r}$ we come to the formulae for the total magnetizations:
\begin{align}
-\frac{\partial\bm{M}_{N}}{\partial\mu_{0}} & =\frac{\beta_{0}}{2\mathrm{i}}\left.\bm{\nabla}_{\bm{q}}\times\left\langle \hat{n}_{\bm{-q}};\hat{\bm{J}}_{N,\bm{q}}\right\rangle _{0}\right|_{\bm{q}\rightarrow0}\,,\\
\bm{M}_{N}-T_{0}\frac{\partial\bm{M}_{N}}{\partial T_{0}} & =\frac{\beta_{0}}{2\mathrm{i}}\left.\bm{\nabla}_{\bm{q}}\times\left\langle \hat{K}_{\bm{-q}};\hat{\bm{J}}_{N,\bm{q}}\right\rangle _{0}\right|_{\bm{q}\rightarrow0}\,,\\
-\frac{\partial\bm{M}_{Q}}{\partial\mu_{0}} & =\frac{\beta_{0}}{2\mathrm{i}}\left.\bm{\nabla}_{\bm{q}}\times\left\langle \hat{n}_{\bm{-q}};\hat{\bm{J}}_{Q,\bm{q}}\right\rangle _{0}\right|_{\bm{q}\rightarrow0}\,,\\
2\bm{M}_{Q}-T_{0}\frac{\partial\bm{M}_{Q}}{\partial T_{0}} & =\frac{\beta_{0}}{2\mathrm{i}}\left.\bm{\nabla}_{\bm{q}}\times\left\langle \hat{K}_{\bm{-q}};\hat{\bm{J}}_{Q,\bm{q}}\right\rangle _{0}\right|_{\bm{q}\rightarrow0}\,.
\end{align}
In the derivation, we assume that $\int\mathrm{d}\bm{r}\bm{\nabla}\times\bm{U}_{ij}(\bm{r})=0$.
This is guaranteed because $\bm{\kappa}_{ij}^{\bm{q}}$ is a well
behaved function.

\section{Details of Derivation for Eqs.~(18-19), Local Equilibrium Currents}

Inserting Eq.~(13) into Eq.~(17), we obtain:

\begin{align}
\bm{J}_{i}^{\mathrm{leq}}(\bm{r}) & \approx\bm{J}_{i}^{\mathrm{eq}}(\bm{r})+\sum_{j=1}^{2}\left(\bm{M}_{ij}\left(\bm{r}\right)\times\bm{\nabla}x_{j}\left(\bm{r}\right)+\int\frac{d\bm{q}}{\left(2\pi\right)^{3}}\bm{\nabla}\times\bm{\kappa}_{ij}^{\bm{q}}\left(\bm{r}\right)x_{j\bm{q}}\right)\,.
\end{align}
Because $x_{j\bm{q}}=\int d\bm{r}^{\prime}x_{j}\left(\bm{r}^{\prime}\right)e^{-\mathrm{i}\bm{q}\cdot\bm{r}^{\prime}}$,
we have: 
\begin{align}
\bm{J}_{i}^{\mathrm{leq}}(\bm{r}) & \approx\bm{J}_{i}^{\mathrm{eq}}(\bm{r})+\sum_{j=1}^{2}\left(\bm{M}_{ij}\left(\bm{r}\right)\times\bm{\nabla}x_{j}\left(\bm{r}\right)+\bm{\nabla}\times\int d\bm{r}^{\prime}\bm{\kappa}_{ij}\left(\bm{r},\bm{r}^{\prime}\right)x_{j}\left(\bm{r}^{\prime}\right)\right)\,,\label{eq:JileqKappa}
\end{align}
where $\bm{\kappa}_{ij}\left(\bm{r},\bm{r}^{\prime}\right)=\int d\bm{q}/\left(2\pi\right)^{3}\bm{\kappa}_{ij}^{\bm{q}}\left(\bm{r}\right)e^{-\mathrm{i}\bm{q}\cdot\bm{r}^{\prime}}$.

We can obtain $\bm{J}_{i}^{\mathrm{eq}}\left(\bm{r}\right)$ through
the scaling law. Without $\psi$ and $\phi$, we have: 
\begin{equation}
\bm{J}_{N(E)}^{\mathrm{eq}}(\bm{r})=\bm{\nabla}\times\bm{M}_{N(E)}(\bm{r})\,.
\end{equation}
When $\psi\left(\bm{r}\right)$ and $\phi\left(\bm{r}\right)$ are
present, according to the scaling law in Eq.~(\ref{eq:scaleN}) and~(\ref{eq:scaleE})
we have: 
\begin{align}
\bm{J}_{1}^{\mathrm{eq}}\left(\bm{r}\right) & =\left[1+\psi(\bm{r})\right]\bm{\nabla}\times\bm{M}_{N}(\bm{r})\,,\label{eq:J1eq}\\
\bm{J}_{2}^{\mathrm{eq}}\left(\bm{r}\right) & =\left[1+\psi(\bm{r})\right]^{2}\left[\bm{\nabla}\times\bm{M}_{E}(\bm{r})-\mu(\bm{r})\bm{\nabla}\times\bm{M}_{N}(\bm{r})\right]\,.\label{eq:J2eq}
\end{align}

For $i=1$, inserting Eq.~(\ref{eq:J1eq}) into Eq.~(\ref{eq:JileqKappa}),
\begin{align}
\bm{J}_{1}^{\mathrm{leq}}(\bm{r}) & \approx\left[1+\psi(\bm{r})\right]\bm{\nabla}\times\bm{M}_{N}(\bm{r})-\bm{M}_{N}\left(\bm{r}\right)\times T_{0}\bm{\nabla}\frac{1}{T}+\bm{\nabla}\times\int d\bm{r}^{\prime}\sum_{j=1}^{2}\bm{\kappa}_{1j}\left(\bm{r},\bm{r}^{\prime}\right)x_{j}\left(\bm{r}^{\prime}\right)\,,\\
 & =\bm{\nabla}\times\left(\left[1+\psi(\bm{r})\right]\bm{M}_{N}(\bm{r})\right)-\frac{1}{\beta}\bm{M}_{N}\left(\bm{r}\right)\times\bm{X}_{2}+\bm{\nabla}\times\int d\bm{r}^{\prime}\sum_{j=1}^{2}\bm{\kappa}_{1j}\left(\bm{r},\bm{r}^{\prime}\right)x_{j}\left(\bm{r}^{\prime}\right)\,,
\end{align}
so we can write: 
\begin{equation}
\bm{J}_{1}^{\mathrm{leq}}(\bm{r})\approx\bm{\nabla}\times\bm{M}_{N}^{\phi,\psi}(\bm{r})-\frac{1}{\beta}\bm{M}_{N}(\bm{r})\times\bm{X}_{2}\,,\label{eq:J1leq}
\end{equation}
where $\bm{M}_{N}^{\phi,\psi}(\bm{r})\equiv\left[1+\psi(\bm{r})\right]\bm{M}_{N}\left(\bm{r},T_{0},\mu_{0}\right)+\delta\bm{M}_{N}(\bm{r})$
and $\delta\bm{M}_{N}\left(\bm{r}\right)\equiv\sum_{j=1}^{2}\int d\bm{r}^{\prime}\bm{\kappa}_{1j}\left(\bm{r},\bm{r}^{\prime}\right)x_{j}\left(\bm{r}^{\prime}\right)$.

Similarly, for $i=2$, inserting Eq.~(\ref{eq:J2eq}) into Eq.~(\ref{eq:JileqKappa}),

\begin{align}
\bm{J}_{2}^{\mathrm{leq}}(\bm{r})\approx & \left[1+\psi(\bm{r})\right]^{2}\left[\bm{\nabla}\times\bm{M}_{E}(\bm{r})-\mu(\bm{r})\bm{\nabla}\times\bm{M}_{N}(\bm{r})\right]+\bm{M}_{N}\left(\bm{r}\right)\times\bm{\nabla}\mu-2\bm{M}_{Q}\times T_{0}\bm{\nabla}\frac{1}{T}\\
 & +\bm{\nabla}\times\int d\bm{r}^{\prime}\sum_{j=1}^{2}\bm{\kappa}_{2j}\left(\bm{r},\bm{r}^{\prime}\right)x_{j}\left(\bm{r}^{\prime}\right)\,,\\
= & \bm{\nabla}\times\left[\left(1+\psi\left(\bm{r}\right)\right)^{2}\left(\bm{M}_{E}(\bm{r})+\phi\left(\bm{r}\right)\bm{M}_{N}\left(\bm{r}\right)\right)\right]-\alpha(\bm{r})\bm{\nabla}\times\left(\left(1+\psi\left(\bm{r}\right)\right)\bm{M}_{N}(\bm{r})\right)\label{eq:MNpsiphi}\\
 & -\frac{1}{\beta}\bm{M}_{N}(\bm{r})\times\bm{X}_{1}-\frac{2}{\beta}\bm{M}_{Q}(\bm{r})\times\bm{X}_{2}+\bm{\nabla}\times\sum_{j=1}^{2}\int d\bm{r}^{\prime}\bm{\kappa}_{2j}\left(\bm{r},\bm{r}^{\prime}\right)x_{j}\left(\bm{r}^{\prime}\right)\,.
\end{align}
Further, by substituting $\bm{M}_{N}^{\phi,\psi}\left(\bm{r}\right)$
into Eq.~(\ref{eq:MNpsiphi}) we can write $\bm{J}_{2}^{\mathrm{leq}}\left(\bm{r}\right)$
as: 
\begin{align}
\bm{J}_{2}^{\mathrm{leq}}(\bm{r})\approx & \bm{\nabla}\times\bm{M}_{E}^{\phi,\psi}(\bm{r})-\alpha(\bm{r})\bm{\nabla}\times\bm{M}_{N}^{\phi,\psi}(\bm{r})-\frac{1}{\beta}\bm{M}_{N}(\bm{r})\times\bm{X}_{1}-\frac{2}{\beta}\bm{M}_{Q}(\bm{r})\times\bm{X}_{2}\,,\label{eq:J2leq}
\end{align}
where $\bm{M}_{E}^{\phi,\psi}\left(\bm{r}\right)\equiv\left(1+\psi\left(\bm{r}\right)\right)^{2}\left(\bm{M}_{E}\left(\bm{r},T_{0},\mu_{0}\right)+\phi\left(\bm{r}\right)\bm{M}_{N}\left(\bm{r},T_{0},\mu_{0}\right)\right)+\delta\bm{M}_{E}\left(\bm{r}\right)$,
$\delta\bm{M}_{E}\left(\bm{r}\right)\equiv\sum_{j=1}^{2}\int d\bm{r}^{\prime}\bm{\kappa}_{2j}^{\prime}\left(\bm{r},\bm{r}^{\prime}\right)x_{j}\left(\bm{r}^{\prime}\right)$,
and $\bm{\kappa}_{2j}^{\prime}\equiv\bm{\kappa}_{2j}+\mu_{0}\bm{\kappa}_{1j}$.

\section{Details of Derivation for Eq.~(22), Definition of Energy Current}

The energy density can be written as:

\begin{equation}
\hat{h}^{\phi,\psi}(\bm{r})=\left[1+\psi(\bm{r})\right]\left\{ \frac{m}{2}\left[\hat{\bm{v}}\hat{\varphi}(\bm{r})\right]^{\dagger}\cdot\left[\hat{\bm{v}}\hat{\varphi}(\bm{r})\right]+\hat{\varphi}^{\dagger}(\bm{r})\left[V(\bm{r})+\phi(\bm{r})\right]\hat{\varphi}(\bm{r})\right\} \,.
\end{equation}
The Schrödinger equation for the system is $\mathrm{i}\hbar\frac{\partial\hat{\varphi}}{\partial t}=\hat{\mathcal{H}}^{\phi,\psi}\hat{\varphi}$
, where $\hat{\mathcal{H}}^{\phi,\psi}\equiv\frac{m}{2}\hat{\bm{v}}\cdot\left[1+\psi(\bm{r})\right]\hat{\bm{v}}+\left[1+\psi(\bm{r})\right]\left[V(\bm{r})+\phi(\bm{r})\right]$.
Therefore, we have: 
\begin{align}
\frac{\partial\hat{h}^{\phi,\psi}(\bm{r})}{\partial t} & =\frac{1}{\mathrm{i}\hbar}\left[1+\psi(\bm{r})\right]\left\{ \frac{m}{2}\left[\hat{\bm{v}}\hat{\varphi}(\bm{r})\right]^{\dagger}\cdot\left[\hat{\bm{v}}\hat{\mathcal{H}}^{\phi,\psi}\hat{\varphi}(\bm{r})\right]-\frac{m}{2}\left[\hat{\bm{v}}\hat{\mathcal{H}}^{\phi,\psi}\hat{\varphi}(\bm{r})\right]^{\dagger}\cdot\left[\hat{\bm{v}}\hat{\varphi}(\bm{r})\right]\right.\\
 & \left.+\hat{\varphi}^{\dagger}(\bm{r})\left[V(\bm{r})+\phi(\bm{r})\right]\left[\hat{\mathcal{H}}^{\phi,\psi}\hat{\varphi}(\bm{r})\right]-\left[\hat{\mathcal{H}}^{\phi,\psi}\hat{\varphi}(\bm{r})\right]^{\dagger}\left[V(\bm{r})+\phi(\bm{r})\right]\hat{\varphi}(\bm{r})\right\} \,,\\
 & =-\bm{\nabla}\cdot\left\{ \frac{1}{2}\left[1+\psi(\bm{r})\right]\left(\left[\hat{\bm{v}}\hat{\varphi}(\bm{r})\right]^{\dagger}\left[\hat{\mathcal{H}}^{\phi,\psi}\hat{\varphi}(\bm{r})\right]+\left[\hat{\mathcal{H}}^{\phi,\psi}\hat{\varphi}(\bm{r})\right]^{\dagger}\left[\hat{\bm{v}}\hat{\varphi}(\bm{r})\right]\right)\right\} \,,
\end{align}
so we can identify $\hat{\bm{J}}_{E}^{\phi,\psi}\left(\bm{r}\right)$
as:
\begin{equation}
\hat{\bm{J}}_{E}^{\phi,\psi}\left(\bm{r}\right)=\frac{1}{2}\left[1+\psi(\bm{r})\right]\left\{ \left[\hat{\bm{v}}\hat{\varphi}(\bm{r})\right]^{\dagger}\left[\hat{\mathcal{H}}^{\phi,\psi}\hat{\varphi}(\bm{r})\right]+\left[\hat{\mathcal{H}}^{\phi,\psi}\hat{\varphi}(\bm{r})\right]^{\dagger}\left[\hat{\bm{v}}\hat{\varphi}(\bm{r})\right]\right\} \,.\label{eq:JE}
\end{equation}

Because: 
\begin{equation}
\hat{\mathcal{H}}^{\phi,\psi}=\left[1+\psi(\bm{r})\right]\left[\hat{\mathcal{H}}_{0}+\phi(\bm{r})\right]-\frac{\mathrm{i}\hbar}{2}\left[\bm{\nabla}\psi(\bm{r})\right]\cdot\hat{\bm{v}}\,,
\end{equation}
where $\hat{\mathcal{H}}_{0}\equiv\hat{\mathcal{H}}^{\phi=0,\psi=0}$,
we obtain the following scaling equation: 
\begin{equation}
\hat{\bm{J}}_{E}^{\phi,\psi}\left(\bm{r}\right)=\left[1+\psi(\bm{r})\right]^{2}\left[\hat{\bm{J}}_{E}\left(\bm{r}\right)+\phi\left(\bm{r}\right)\hat{\bm{J}}_{N}\left(\bm{r}\right)\right]+\bm{\nabla}\left(1+\psi(\bm{r})\right)^{2}\times\hat{\bm{\Lambda}}\left(\bm{r}\right)\,,\label{eq:jescale}
\end{equation}
where $\hat{\bm{\Lambda}}\left(\bm{r}\right)=\frac{\hbar}{8\mathrm{i}}\left(\hat{\bm{v}}\hat{\varphi}\right)^{\dagger}\times\left(\hat{\bm{v}}\hat{\varphi}\right)$. 

In order to satisfy the scaling law Eq.~(5), we redefine the energy
current operator as:
\begin{eqnarray}
\hat{\bm{J}}_{E}^{\phi,\psi}\left(\bm{r}\right) & \rightarrow & \hat{\bm{J}}_{E}^{\phi,\psi}\left(\bm{r}\right)-\bm{\nabla}\times\left(\left(1+\psi(\bm{r})\right)^{2}\hat{\bm{\Lambda}}\left(\bm{r}\right)\right)\,,\\
\hat{\bm{J}}_{E}(\bm{r}) & \rightarrow & \hat{\bm{J}}_{E}\left(\bm{r}\right)-\bm{\nabla}\times\hat{\bm{\Lambda}}\left(\bm{r}\right)\,.
\end{eqnarray}
This is exactly the energy current definition Eq.~(22). It is straightforward
to show that modified energy current operator satisfies the scaling
law Eq.~(5).

The particle current operator is defined as usual. It automatically
satisfies the corresponding scaling law Eq.~(4).

\section{Details of Derivation for Eq.~(23), Kubo Contribution}

The thermal current operator $\hat{J}_{Qx}\left(\bm{r}\right)$ is:
\begin{equation}
\hat{J}_{Qx}\left(\bm{r}\right)=\frac{\left(\hat{K}\hat{\varphi}\left(\bm{r}\right)\right)^{\dagger}\hat{v}_{x}\hat{\varphi}\left(\bm{r}\right)+\left(\hat{v}_{x}\hat{\varphi}\left(\bm{r}\right)\right)^{\dagger}\hat{K}\hat{\varphi}\left(\bm{r}\right)}{2}-\frac{\hbar}{8\mathrm{i}}\sum_{\gamma}\nabla_{\gamma}\left(\left(\hat{v}_{x}\hat{\varphi}\left(\bm{r}\right)\right)^{\dagger}\hat{v}_{\gamma}\hat{\varphi}\left(\bm{r}\right)-\left(\hat{v}_{\gamma}\hat{\varphi}\left(\bm{r}\right)\right)^{\dagger}\hat{v}_{x}\hat{\varphi}\left(\bm{r}\right)\right)\,,\label{eq:thermalcurrent}
\end{equation}
where $\gamma=x,y,z$ and we have set $\phi\left(\bm{r}\right)=0$
and $\psi\left(\bm{r}\right)=0$. According to our definition, we
have~\cite{Mahan2000} $\kappa_{xy}^{\mathrm{Kubo}}\equiv\frac{L_{xy}^{(22)}}{k_{B}T_{0}^{2}}$
and:

\begin{align}
L_{xy}^{(22)}= & \frac{1}{V}\int_{0}^{\infty}dte^{-st}\left\langle \hat{J}_{Qy};\,\hat{J}_{Qx}\left(t\right)\right\rangle _{0}\,,\\
= & -\frac{\hbar}{\beta_{0}V}\sum_{n\bm{k},n^{\prime}\bm{k}^{\prime}}\frac{f_{n\bm{k}}-f_{n^{\prime}\bm{k}^{\prime}}}{\mathrm{i}\left(\epsilon_{n\bm{k}}-\epsilon_{n^{\prime}\bm{k}^{\prime}}\right)^{2}}\left\langle \psi_{n\bm{k}}\right|\hat{\mathcal{J}}_{Qy}\left|\psi_{n^{\prime}\bm{k}^{\prime}}\right\rangle \left\langle \psi_{n^{\prime}\bm{k}^{\prime}}\right|\hat{\mathcal{J}}_{Qx}\left|\psi_{n\bm{k}}\right\rangle \,,
\end{align}
where $\psi_{n\bm{k}}$ is the Bloch wave function for band $n$ and
quasi-momentum $\bm{k}$, $f_{n\bm{k}}\equiv f(\epsilon_{n\bm{k}})$
is the Fermi distribution function, and $\epsilon_{n\bm{k}}$ is the
electron dispersion. According to our definition Eq.~(\ref{eq:thermalcurrent})
for $\hat{J}_{Qx}$, we have: 
\begin{align}
\left\langle \psi_{n^{\prime}\bm{k}^{\prime}}\right|\hat{\mathcal{J}}_{Qx}\left|\psi_{n\bm{k}}\right\rangle = & \frac{\left\langle \hat{K}\psi_{n^{\prime}\bm{k}^{\prime}}\left|\hat{v}_{x}\psi_{n\bm{k}}\right\rangle \right.+\left\langle \hat{v}_{x}\psi_{n^{\prime}\bm{k}^{\prime}}\left|\hat{K}\psi_{n\bm{k}}\right\rangle \right.}{2}-\frac{\hbar}{8\mathrm{i}}\sum_{\gamma}\left[\left\langle \nabla_{\gamma}\hat{v}_{x}\psi_{n^{\prime}\bm{k}^{\prime}}\left|\hat{v}_{\gamma}\psi_{n\bm{k}}\right\rangle \right.+\left\langle \hat{v}_{x}\psi_{n^{\prime}\bm{k}^{\prime}}\left|\nabla_{\gamma}\hat{v}_{\gamma}\psi_{n\bm{k}}\right\rangle \right.\right.\\
 & \left.-\left\langle \nabla_{\gamma}\hat{v}_{\gamma}\psi_{n^{\prime}\bm{k}^{\prime}}\left|\hat{v}_{x}\psi_{n\bm{k}}\right\rangle \right.-\left\langle \hat{v}_{\gamma}\psi_{n^{\prime}\bm{k}^{\prime}}\left|\nabla_{\gamma}\hat{v}_{x}\psi_{n\bm{k}}\right\rangle \right.\right]\,.
\end{align}
Note: 
\begin{align}
\left\langle \hat{K}\psi_{n^{\prime}\bm{k}^{\prime}}\left|\hat{v}_{x}\psi_{n\bm{k}}\right\rangle \right. & =\left(\epsilon_{n^{\prime}\bm{k}}-\mu_{0}\right)\left\langle u_{n^{\prime}\bm{k}}\right|\hat{v}_{\bm{k}x}\left|u_{n\bm{k}}\right\rangle \delta_{\bm{k}\bm{k}^{\prime}}\,,
\end{align}
with $\hat{v}_{\bm{k}x}=\partial\hat{\mathcal{H}}_{\bm{k}}/\partial\left(\hbar k_{x}\right)$,
and:
\begin{align}
\left\langle \nabla_{\gamma}\hat{v}_{x}\psi_{n^{\prime}\bm{k}^{\prime}}\left|\hat{v}_{\gamma}\psi_{n\bm{k}}\right\rangle \right.+\left\langle \hat{v}_{x}\psi_{n^{\prime}\bm{k}^{\prime}}\left|\nabla_{\gamma}\hat{v}_{\gamma}\psi_{n\bm{k}}\right\rangle \right. & =-\left\langle \psi_{n^{\prime}\bm{k}^{\prime}}\right|\hat{v}_{x}\nabla_{\gamma}\hat{v}_{\gamma}\left|\psi_{n\bm{k}}\right\rangle +\left\langle \psi_{n^{\prime}\bm{k}^{\prime}}\right|\hat{v}_{x}\nabla_{\gamma}\hat{v}_{\gamma}\left|\psi_{n\bm{k}}\right\rangle \,,\\
 & =0\,,
\end{align}
and similarly, $\left\langle \nabla_{\gamma}\hat{v}_{\gamma}\psi_{n^{\prime}\bm{k}^{\prime}}\left|\hat{v}_{x}\psi_{n\bm{k}}\right\rangle \right.+\left\langle \hat{v}_{\gamma}\psi_{n^{\prime}\bm{k}^{\prime}}\left|\nabla_{\gamma}\hat{v}_{x}\psi_{n\bm{k}}\right\rangle \right.=0$,
so we come to: 
\begin{align}
L_{xy}^{(22)}= & -\frac{\hbar}{\beta_{0}V}\sum_{n\neq n^{\prime}\bm{k}}\frac{\left(f_{n\bm{k}}-f_{n^{\prime}\bm{k}}\right)\left(\epsilon_{n\bm{k}}+\epsilon_{n^{\prime}\bm{k}}-2\mu_{0}\right)^{2}}{4\mathrm{i}\left(\epsilon_{n\bm{k}}-\epsilon_{n^{\prime}\bm{k}}\right)^{2}}\left\langle u_{n\bm{k}}\right|\hat{v}_{\bm{k}y}\left|u_{n^{\prime}\bm{k}}\right\rangle \left\langle \psi_{n^{\prime}\bm{k}}\right|\hat{v}_{\bm{k}x}\left|\psi_{n\bm{k}}\right\rangle \,,\\
= & -\frac{\hbar}{2\beta_{0}V}\sum_{n\neq n^{\prime}\bm{k}}\frac{f_{n\bm{k}}\left(\epsilon_{n\bm{k}}+\epsilon_{n^{\prime}\bm{k}}-2\mu_{0}\right)^{2}}{\left(\epsilon_{n\bm{k}}-\epsilon_{n^{\prime}\bm{k}}\right)^{2}}\mathrm{Im}\left[\left\langle u_{n\bm{k}}\right|\hat{v}_{\bm{k}y}\left|u_{n^{\prime}\bm{k}}\right\rangle \left\langle u_{n^{\prime}\bm{k}}\right|\hat{v}_{\bm{k}x}\left|u_{n\bm{k}}\right\rangle \right]\,.
\end{align}
Using the identity:
\begin{equation}
\left\langle u_{n^{\prime}\bm{k}}\right|\hat{v}_{\bm{k}x}\left|u_{n\bm{k}}\right\rangle =\frac{1}{\hbar}\left(\epsilon_{n\bm{k}}-\epsilon_{n^{\prime}\bm{k}}\right)\left\langle u_{n^{\prime}\bm{k}}\left|\frac{\partial u_{n\bm{k}}}{\partial k_{x}}\right\rangle \right.\,,\label{eq:identity}
\end{equation}
we have:
\begin{equation}
L_{xy}^{\left(22\right)}=\frac{1}{2\beta_{0}\hbar V}\sum_{n\bm{k}}\mathrm{Im}\left[\left\langle \frac{\partial u_{n\bm{k}}}{\partial k_{x}}\right|\left(\hat{\mathcal{H}}_{\bm{k}}+\epsilon_{n\bm{k}}-2\mu_{0}\right)^{2}\left|\frac{\partial u_{n\bm{k}}}{\partial k_{y}}\right\rangle \right]f_{n\bm{k}}\,.
\end{equation}
The formula can be rewritten as the alternative form. We introduce
the new notations:
\begin{align}
m_{2}\left(\epsilon\right)\equiv & \frac{1}{\hbar}\mathrm{Im}\sum_{n\bm{k}}\left\langle \frac{\partial u_{n\bm{k}}}{\partial k_{x}}\right|\left(\hat{\mathcal{H}}_{\bm{k}}-\epsilon\right)^{2}\left|\frac{\partial u_{n\bm{k}}}{\partial k_{y}}\right\rangle \delta\left(\epsilon-\epsilon_{n\bm{k}}\right)\,,\label{eq:m2}\\
m_{1}\left(\epsilon\right)\equiv & \frac{1}{\hbar}\mathrm{Im}\sum_{n\bm{k}}\left\langle \frac{\partial u_{n\bm{k}}}{\partial k_{x}}\right|\left(\hat{\mathcal{H}}_{\bm{k}}-\epsilon\right)\left|\frac{\partial u_{n\bm{k}}}{\partial k_{y}}\right\rangle \delta\left(\epsilon-\epsilon_{n\bm{k}}\right)\,,\label{eq:m1}\\
\Omega_{z}\left(\epsilon\right)\equiv & -\frac{2}{\hbar}\mathrm{Im}\sum_{n\bm{k}}\left.\left\langle \frac{\partial u_{n\bm{k}}}{\partial k_{x}}\right|\frac{\partial u_{n\bm{k}}}{\partial k_{y}}\right\rangle \delta\left(\epsilon-\epsilon_{n\bm{k}}\right)\,.\label{eq:omega}
\end{align}
Therefore, we can express $\kappa_{xy}^{\mathrm{Kubo}}$ as: 
\begin{equation}
\kappa_{xy}^{\mathrm{Kubo}}=\frac{1}{2T_{0}V}\int d\epsilon\left[m_{2}\left(\epsilon\right)+4\left(\epsilon-\mu_{0}\right)m_{1}\left(\epsilon\right)-2\left(\epsilon-\mu_{0}\right)^{2}\Omega_{z}\left(\epsilon\right)\right]f\left(\epsilon\right)\,.
\end{equation}
It is easy to see $\kappa_{xy}^{\mathrm{Kubo}}$ is divergent in the
low temperature limit.

\section{Details of Derivation for Eq.~(24), Energy Magnetization}

To calculate $M_{Q,z}$, we use $2\bm{M}_{Q}-T_{0}\frac{\partial\bm{M}_{Q}}{\partial T_{0}}=\frac{\beta_{0}}{2\mathrm{i}}\left.\bm{\nabla}_{\bm{q}}\times\left\langle \hat{K}_{\bm{-q}};\hat{\bm{J}}_{Q,\bm{q}}\right\rangle _{0}\right|_{\bm{q}\rightarrow0}$.
We can show: 
\begin{equation}
\tilde{M}_{Q,z}\text{\ensuremath{\equiv}}\frac{\beta_{0}}{2\mathrm{i}}\left.\bm{\nabla}_{\bm{q}}\times\left\langle \hat{K}_{\bm{-q}};\hat{\bm{J}}_{Q,\bm{q}}\right\rangle _{0}\right|_{z,\bm{q}\rightarrow0}=-\beta_{0}\left.\frac{\partial}{\mathrm{i}\partial q_{y}}\left\langle \hat{K}_{\bm{-q}};\hat{J}_{Qx,\bm{q}}\right\rangle _{0}\right|_{\bm{q}\rightarrow0}\,.
\end{equation}
So we have: 
\begin{equation}
\tilde{M}_{Q,z}=\frac{\partial}{\mathrm{i}\partial q_{y}}\sum_{n\bm{k},n^{\prime}\bm{k}^{\prime}}\frac{f_{n\bm{k}}-f_{n^{\prime}\bm{k}^{\prime}}}{\epsilon_{n\bm{k}}-\epsilon_{n^{\prime}\bm{k}^{\prime}}}\left\langle \psi_{n\bm{k}}\right|\frac{\hat{K}e^{\mathrm{i}\bm{q}\cdot\bm{r}}+e^{\mathrm{i}\bm{q}\cdot\bm{r}}\hat{K}}{2}\left|\psi_{n^{\prime}\bm{k}^{\prime}}\right\rangle \left\langle \psi_{n^{\prime}\bm{k}^{\prime}}\right|\hat{\mathcal{J}}_{Qx,\bm{q}}\left|\psi_{n\bm{k}}\right\rangle \,.
\end{equation}
We have a careful calculation of $\left\langle \psi_{n^{\prime}\bm{k}^{\prime}}\right|\hat{\mathcal{J}}_{Qx,\bm{q}}\left|\psi_{n\bm{k}}\right\rangle $,

\begin{align}
\left\langle \psi_{n^{\prime}\bm{k}^{\prime}}\right|\hat{\mathcal{J}}_{Qx,\bm{q}}\left|\psi_{n\bm{k}}\right\rangle = & \frac{\left\langle \hat{K}\psi_{n^{\prime}\bm{k}^{\prime}}\right|e^{-\mathrm{i}\bm{q}\cdot\bm{r}}\left|\hat{v}_{x}\psi_{n\bm{k}}\right\rangle +\left\langle \hat{v}_{x}\psi_{n^{\prime}\bm{k}^{\prime}}\right|e^{-\mathrm{i}\bm{q}\cdot\bm{r}}\left|\hat{K}\psi_{n\bm{k}}\right\rangle }{2}\label{eq:Kv}\\
 & -\frac{\hbar}{8\mathrm{i}}\sum_{\gamma}\left[\left\langle \nabla_{\gamma}\hat{v}_{x}\psi_{n^{\prime}\bm{k}^{\prime}}\right|e^{-\mathrm{i}\bm{q}\cdot\bm{r}}\left|\hat{v}_{\gamma}\psi_{n\bm{k}}\right\rangle +\left\langle \hat{v}_{x}\psi_{n^{\prime}\bm{k}^{\prime}}\right|e^{-\mathrm{i}\bm{q}\cdot\bm{r}}\left|\nabla_{\gamma}\hat{v}_{\gamma}\psi_{n\bm{k}}\right\rangle \right.\label{eq:vxvgamma}\\
 & \left.-\left\langle \nabla_{\gamma}\hat{v}_{\gamma}\psi_{n^{\prime}\bm{k}^{\prime}}\right|e^{-\mathrm{i}\bm{q}\cdot\bm{r}}\left|\hat{v}_{x}\psi_{n\bm{k}}\right\rangle -\left\langle \hat{v}_{\gamma}\psi_{n^{\prime}\bm{k}^{\prime}}\right|e^{-\mathrm{i}\bm{q}\cdot\bm{r}}\left|\nabla_{\gamma}\hat{v}_{x}\psi_{n\bm{k}}\right\rangle \right]\,.\label{eq:vgammavx}
\end{align}
In Eq.~(\ref{eq:Kv}), we can show:
\begin{align}
\left\langle \hat{K}\psi_{n^{\prime}\bm{k}^{\prime}}\right|e^{-\mathrm{i}\bm{q}\cdot\bm{r}}\left|\hat{v}_{x}\psi_{n\bm{k}}\right\rangle  & =\left\langle u_{n^{\prime}\bm{k}^{\prime}}\right|e^{-\mathrm{i}\bm{k}^{\prime}\cdot\bm{r}}\hat{K}e^{-\mathrm{i}\bm{q}\cdot\bm{r}}\hat{v}_{x}e^{\mathrm{i}\bm{k}\cdot\bm{r}}\left|u_{n\bm{k}}\right\rangle \,,\\
 & =\left\langle u_{n^{\prime}\bm{k}-\bm{q}}\right|\hat{K}_{\bm{k}-\bm{q}}\hat{v}_{\bm{k}x}\left|u_{n\bm{k}}\right\rangle \delta_{\bm{k}^{\prime},\bm{k}-\bm{q}}\,.
\end{align}
In Eq.~(\ref{eq:vxvgamma}), similarly:
\begin{align}
\left\langle \nabla_{\gamma}\hat{v}_{x}\psi_{n^{\prime}\bm{k}^{\prime}}\right|e^{-\mathrm{i}\bm{q}\cdot\bm{r}}\left|\hat{v}_{\gamma}\psi_{n\bm{k}}\right\rangle  & =-\left\langle u_{n^{\prime}\bm{k}^{\prime}}\right|e^{-\mathrm{i}\bm{k}^{\prime}\cdot\bm{r}}\hat{v}_{x}\nabla_{\gamma}e^{-\mathrm{i}\bm{q}\cdot\bm{r}}\hat{v}_{\gamma}e^{\mathrm{i}\bm{k}\cdot\bm{r}}\left|u_{n\bm{k}}\right\rangle \,,\\
 & =-\left\langle u_{n^{\prime}\bm{k}-\bm{q}}\right|\hat{v}_{\bm{k}-\bm{q}x}\left(\nabla_{\gamma}+\mathrm{i}k_{\gamma}-\mathrm{i}q_{\gamma}\right)\hat{v}_{\bm{k}\gamma}\left|u_{n\bm{k}}\right\rangle \delta_{\bm{k}^{\prime},\bm{k}-\bm{q}}\,.
\end{align}
and: 
\begin{align}
\left\langle \hat{v}_{x}\psi_{n^{\prime}\bm{k}^{\prime}}\right|e^{-\mathrm{i}\bm{q}\cdot\bm{r}}\left|\nabla_{\gamma}\hat{v}_{\gamma}\psi_{n\bm{k}}\right\rangle  & =\left\langle u_{n^{\prime}\bm{k}-\bm{q}}\right|\hat{v}_{\bm{k}-\bm{q}x}\left(\nabla_{\gamma}+\mathrm{i}k_{\gamma}\right)\hat{v}_{\bm{k}\gamma}\left|u_{n\bm{k}}\right\rangle \delta_{\bm{k}^{\prime},\bm{k}-\bm{q}}\,,\\
\left\langle \nabla_{\gamma}\hat{v}_{\gamma}\psi_{n^{\prime}\bm{k}^{\prime}}\right|e^{-\mathrm{i}\bm{q}\cdot\bm{r}}\left|\hat{v}_{x}\psi_{n\bm{k}}\right\rangle  & =-\left\langle u_{n^{\prime}\bm{k}-\bm{q}}\right|\hat{v}_{\bm{k}-\bm{q}\gamma}\left(\nabla_{\gamma}+\mathrm{i}k_{\gamma}-\mathrm{i}q_{\gamma}\right)\hat{v}_{\bm{k}x}\left|u_{n\bm{k}}\right\rangle \delta_{\bm{k}^{\prime},\bm{k}-\bm{q}}\,,\\
\left\langle \hat{v}_{\gamma}\psi_{n^{\prime}\bm{k}^{\prime}}\right|e^{-\mathrm{i}\bm{q}\cdot\bm{r}}\left|\nabla_{\gamma}\hat{v}_{x}\psi_{n\bm{k}}\right\rangle  & =\left\langle u_{n^{\prime}\bm{k}-\bm{q}}\right|\hat{v}_{\bm{k}-\bm{q}\gamma}\left(\nabla_{\gamma}+\mathrm{i}k_{\gamma}\right)\hat{v}_{\bm{k}x}\left|u_{n\bm{k}}\right\rangle \delta_{\bm{k}^{\prime},\bm{k}-\bm{q}}\,.
\end{align}
Therefore, $\left\langle \psi_{n^{\prime}\bm{k}^{\prime}}\right|\hat{J}_{Qx}e^{-\mathrm{i}\bm{q}\cdot\bm{r}}\left|\psi_{n\bm{k}}\right\rangle $
is,
\begin{equation}
\left\langle \psi_{n^{\prime}\bm{k}^{\prime}}\right|\hat{J}_{Qx}e^{-\mathrm{i}\bm{q}\cdot\bm{r}}\left|\psi_{n\bm{k}}\right\rangle =\left\langle u_{n^{\prime}\bm{k}-\bm{q}}\right|\frac{\hat{K}_{\bm{k}-\bm{q}}\hat{v}_{\bm{k}x}+\hat{v}_{\bm{k}-\bm{q}x}\hat{K}_{\bm{k}}}{2}-\frac{\sum_{\gamma}\hbar q_{\gamma}\left(\hat{v}_{\bm{k}-\bm{q}x}\hat{v}_{\bm{k}\gamma}-\hat{v}_{\bm{k}-\bm{q}\gamma}\hat{v}_{\bm{k}x}\right)}{8}\left|u_{n\bm{k}}\right\rangle \delta_{\bm{k}^{\prime},\bm{k}-\bm{q}}\,.
\end{equation}
$\tilde{M}_{Q,z}$ can be simplified as,

\begin{align}
\tilde{M}_{Q,z}= & \frac{\partial}{\mathrm{i}\partial q_{y}}\sum_{n\bm{k},n^{\prime}}\frac{f_{n\bm{k}}-f_{n^{\prime}\bm{k}-\bm{q}}}{\epsilon_{n\bm{k}}-\epsilon_{n^{\prime}\bm{k}-\bm{q}}}\left\langle u_{n\bm{k}}\right|\frac{\hat{K}_{\bm{k}}+\hat{K}_{\bm{k}-\bm{q}}}{2}\left|u_{n^{\prime}\bm{k}-\bm{q}}\right\rangle \left\langle u_{n^{\prime}\bm{k}-\bm{q}}\right|\frac{\hat{K}_{\bm{k}-\bm{q}}\hat{v}_{\bm{k}x}+\hat{v}_{\bm{k}-\bm{q}x}\hat{K}_{\bm{k}}}{2}\\
 & -\frac{\sum_{\gamma}\hbar q_{\gamma}\left(\hat{v}_{\bm{k}-\bm{q}x}\hat{v}_{\bm{k}\gamma}-\hat{v}_{\bm{k}-\bm{q}\gamma}\hat{v}_{\bm{k}x}\right)}{8}\left|u_{n\bm{k}}\right\rangle \,.
\end{align}

First, we calculate $\tilde{M}_{Q,z}^{\mathrm{inter}}$ for $n\neq n^{\prime}$.
When $\bm{q}\rightarrow0$, we have:
\begin{equation}
\tilde{M}_{Q,z}^{\mathrm{inter}}=-\frac{1}{4}\sum_{n\neq n^{\prime}\bm{k}}\left(\epsilon_{n\bm{k}}+\epsilon_{n^{\prime}\bm{k}}-2\mu_{0}\right)^{2}\mathrm{Im}\left[\left\langle u_{n\bm{k}}\left|\frac{\partial u_{n^{\prime}\bm{k}}}{\partial k_{y}}\right\rangle \right.\left\langle u_{n^{\prime}\bm{k}}\right|\hat{v}_{\bm{k}x}\left|u_{n\bm{k}}\right\rangle \right]\frac{f_{n\bm{k}}-f_{n^{\prime}\bm{k}}}{\epsilon_{n\bm{k}}-\epsilon_{n^{\prime}\bm{k}}}\,.
\end{equation}
Using the identity Eq.~(\ref{eq:identity}), we finally come to:
\begin{equation}
\tilde{M}_{Q,z}^{\mathrm{inter}}=-\frac{1}{2\hbar}\sum_{n\bm{k}}\mathrm{Im}\left[\left\langle \frac{\partial u_{n\bm{k}}}{\partial k_{x}}\right|\left(\hat{\mathcal{H}}_{\bm{k}}+\epsilon_{n\bm{k}}-2\mu_{0}\right)^{2}\left|\frac{\partial u_{n\bm{k}}}{\partial k_{y}}\right\rangle \right]f_{n\bm{k}}\,.
\end{equation}

Next, we calculate $\tilde{M}_{Q,z}^{\mathrm{intra}}$ for $n=n^{\prime}$.
When $\bm{q}\rightarrow0$, we have:

\begin{align}
\tilde{M}_{Q,z}^{\mathrm{intra}}= & -\frac{1}{4}\sum_{n\bm{k}}4\left(\epsilon_{n\bm{k}}-\mu_{0}\right)^{2}\mathrm{Im}\left[\left\langle u_{n\bm{k}}\left|\frac{\partial u_{n\bm{k}}}{\partial k_{y}}\right\rangle \right.\left\langle u_{n\bm{k}}\right|\hat{v}_{\bm{k}x}\left|u_{n\bm{k}}\right\rangle \right]f_{n\bm{k}}^{\prime}\\
 & -\frac{1}{4}\sum_{n\bm{k}}2\left(\epsilon_{n\bm{k}}-\mu_{0}\right)\mathrm{Im}\left[\left\langle \frac{\partial u_{n\bm{k}}}{\partial k_{y}}\right|\hat{K}_{\bm{k}}\hat{v}_{\bm{k}x}+\hat{v}_{\bm{k}x}\hat{K}_{\bm{k}}\left|u_{n\bm{k}}\right\rangle \right]f_{n\bm{k}}^{\prime}\\
 & -\frac{\hbar}{4}\sum_{n\bm{k}}\left(\epsilon_{n\bm{k}}-\mu_{0}\right)\mathrm{Im}\left[\left\langle u_{n\bm{k}}\right|\hat{v}_{\bm{k}y}\hat{v}_{\bm{k}x}\left|u_{n\bm{k}}\right\rangle \right]f_{n\bm{k}}^{\prime}\,.
\end{align}
Using the identity Eq.~(\ref{eq:identity}) and after some simple
algebra, we obtain: 
\begin{equation}
\tilde{M}_{Q,z}^{\mathrm{intra}}=-\frac{1}{4\hbar}\sum_{n\bm{k}}\mathrm{Im}\left[\left\langle \frac{\partial u_{n\bm{k}}}{\partial k_{x}}\right|\left(\epsilon_{n\bm{k}}-\hat{\mathcal{H}}_{\bm{k}}\right)^{2}-4\left(\epsilon_{n\bm{k}}-\mu_{0}\right)\left(\epsilon_{n\bm{k}}-\hat{\mathcal{H}}_{\bm{k}}\right)\left|\frac{\partial u_{n\bm{k}}}{\partial k_{y}}\right\rangle \right]\left(\epsilon_{n\bm{k}}-\mu_{0}\right)f_{n\bm{k}}^{\prime}\,.
\end{equation}

Therefore, we have: 
\begin{align}
\tilde{M}_{Q,z}= & -\frac{1}{2\hbar}\sum_{n\bm{k}}\mathrm{Im}\left[\left\langle \frac{\partial u_{n\bm{k}}}{\partial k_{x}}\right|\left(\hat{\mathcal{H}}_{\bm{k}}+\epsilon_{n\bm{k}}-2\mu_{0}\right)^{2}\left|\frac{\partial u_{n\bm{k}}}{\partial k_{y}}\right\rangle \right]f_{n\bm{k}}\\
 & -\frac{1}{4\hbar}\sum_{n\bm{k}}\mathrm{Im}\left[\left\langle \frac{\partial u_{n\bm{k}}}{\partial k_{x}}\right|\left(\epsilon_{n\bm{k}}-\hat{\mathcal{H}}_{\bm{k}}\right)^{2}-4\left(\epsilon_{n\bm{k}}-\mu_{0}\right)\left(\epsilon_{n\bm{k}}-\hat{\mathcal{H}}_{\bm{k}}\right)\left|\frac{\partial u_{n\bm{k}}}{\partial k_{y}}\right\rangle \right]\left(\epsilon_{n\bm{k}}-\mu_{0}\right)f_{n\bm{k}}^{\prime}\,.
\end{align}
We use $2M_{Q,z}-T_{0}(\partial M_{Q,z}/\partial T_{0})=\tilde{M}_{Q,z}$
to obtain $M_{Q,z}$. Using the notations of Eqs.~(\ref{eq:m2})--(\ref{eq:omega}),
we obtain: 
\begin{align}
M_{Q,z}= & -\frac{1}{2}\int d\epsilon\left[\frac{1}{2}m_{2}\left(\epsilon\right)f\left(\epsilon\right)+2\left(\epsilon-\mu_{0}\right)m_{1}\left(\epsilon\right)f\left(\epsilon\right)-2\Omega_{z}\left(\epsilon\right)\int_{0}^{\epsilon-\mu_{0}}dxxf\left(x\right)\right]\,.
\end{align}